\documentclass[journal,twocolumn]{IEEEtran}

\makeatletter
\usepackage{graphicx}
\usepackage{graphics,color}
\usepackage{amsmath}
\usepackage{cite}
\usepackage{url}
\usepackage{xspace}
\usepackage{enumitem}  
\usepackage[bookmarks]{hyperref}
\usepackage{scrextend}

\usepackage{epstopdf}
\usepackage{latexsym}
\usepackage{amssymb}
\usepackage{epsfig, psfrag}
\usepackage{amsthm} 
\usepackage{placeins}

 \DeclareMathOperator*{\argmax}{arg\,max}
 
\newtheorem{theorem}{Theorem}
\newtheorem{lemma}{Lemma}
\newtheorem{proposition}{Proposition}
\newtheorem{corollary}{Corollary}

\newtheorem{fact}{Fact}
\theoremstyle{definition}
\newtheorem{definition} {Definition}
\newtheorem{remarks}{Remark}
\newtheorem{example}{Example}

\newfont{\boldlarge}{msbm10 scaled 1100}

\newcommand{\ignore}[1]{}

\def\expe{\mathbb{E}}   

\def\argmax{\mathop{\rm argmax}}

\def\P{\mathsf{P}}
\def\ber{\mathsf{Ber}}

\newcommand{\add}[1]{{\color{black}{#1}}}

\newcommand{\kl}[2]{D\left( \left. #1 \right\| #2 \right)}

\def\mc{\mathcal}
\def\mbf{\mathbf}
\def\mbb{\mathbb}

\usepackage{mathtools}
\mathtoolsset{showonlyrefs=true}

\renewcommand{\qed}{\nobreak \ifvmode \relax \else
      \ifdim\lastskip<1.5em \hskip-\lastskip
      \hskip1.5em plus0em minus0.5em \fi \nobreak
      \vrule height0.2em width0.5em depth0.4em\fi}
\IEEEoverridecommandlockouts
\begin{document}

\sloppy

\title{On Error Exponents of Almost-Fixed-Length Channel Codes and Hypothesis Tests}


\author{\IEEEauthorblockN{Anusha Lalitha$^1$ and Tara Javidi$^{2}$}\\
\IEEEauthorblockA{$^1$ Stanford University, $^2$ University of California San Diego \\
Email: alalitha@stanford.edu, tjavidi@eng.ucsd.edu}
\thanks{This paper was presented in part in~\cite{7541591}.}  
}

\maketitle

\begin{abstract} 
We examine a new class of channel coding strategies, and hypothesis tests referred to as almost-fixed-length strategies that have little flexibility in the stopping time over fixed-length strategies. The stopping time of these strategies is allowed to be slightly large only on a rare set of sample paths with an exponentially small probability. We show that almost-fixed-length channel coding strategies can achieve Burnashev's optimal error exponent. Similarly, almost-fixed length hypothesis tests are shown to bridge the gap between hypothesis testing with fixed sample size and sequential hypothesis testing and improve the trade-off between type-I and type-II error exponents.
\end{abstract}


\section{Introduction}

It is well known that the capacity of a discrete memoryless channel (DMC) does not increase in the presence of a feedback channel~\cite{CoverBook2nd}. Nevertheless, for a fixed information transmission rate, feedback can reduce the complexity of the encoder and the decoder, and reduce the probability of error. For example, for the binary erasure channel (BEC) with feedback, it is possible to implement a simple and low complexity communication strategy that achieves capacity and provides zero probability of error for all rates below capacity.  The strategy is to send each information bit repeatedly until it is received unerased. This strategy has a random stopping time that depends on how often the channel erased the information bits.  If we restrict ourselves to a fixed-length communication strategy, we can no longer guarantee a zero probability of error for any rate below capacity.  In general, for symmetric DMCs \footnote{Here, we consider symmetric DMCs channels with critical rate strictly below capacity.}, for rates close to capacity, having feedback does not improve the error exponent for fixed-length strategies but provides an improvement for variable-length strategies. We see that the improvement in error exponent when feedback is available not only comes from the encoder's ability to observe how well the decoder is doing but also because the encoder can adapt the length of the code. This implies the variability in the stopping time, i.e., the capability to adapt is fundamental to the boost in error exponent.  A question that naturally arises is: how much variability is needed in the stopping time of channel codes with feedback to see an improvement over the error exponent of fixed-length codes? 

Traditionally, variable-length channel codes are studied under average length constraints designed so that the expected stopping time is bounded. For the variable-length codes, Burnashev~\cite{Burnashev76} provided an exact expression for the maximum achievable error exponent for all DMCs with feedback. Burnashev's optimal error exponent exceeds the upper bound on the maximum achievable error exponent for any fixed-length code with feedback, i.e., the sphere packing bound. However, in many practical applications that require small or moderate latency, using strategies that satisfy only average-length constraints for smaller error probability has significant limitations. Variable-length codes do not prohibit the stopping time from occasionally being very large and do not limit the stopping time variability. The main contribution of this paper is to demonstrate that the variability in the stopping time of channel codes with feedback need not be significant to improve the error exponents. Specifically, we answer the question we raised by showing that Burnashev's optimal error exponent can be achieved with codes whose stopping time is kept fixed for almost all sample paths except for an exponentially rare set. 

This paper introduces a new class of channel coding strategies referred to as \emph{almost-fixed-length channel codes} in which the stopping time is less than $\ell$ with high probability and with an exponentially small probability the stopping time is allowed exceed $\ell$ but remains bounded by $K\ell$ where $K$ is a constant. Furthermore, the probability of stopping time exceeding $\ell$ approaches zero exponentially fast with an exponent $\gamma > 0$. This implies that the variance of the stopping time of almost-fixed-length channel codes approaches zero as $\ell$ grows. Hence, the variability of the stopping time for almost-fixed-length channel codes is significantly more restricted than that of variable-length channel codes. This paper shows that it is possible to achieve optimal error exponent of variable-length strategies using almost-fixed-length strategies. Hence, growing variability is not essential to achieve optimal error exponents. The following example shows that significant gap in the performance of the two settings.

\begin{figure}[!htb]
    \includegraphics[width=0.45\textwidth]{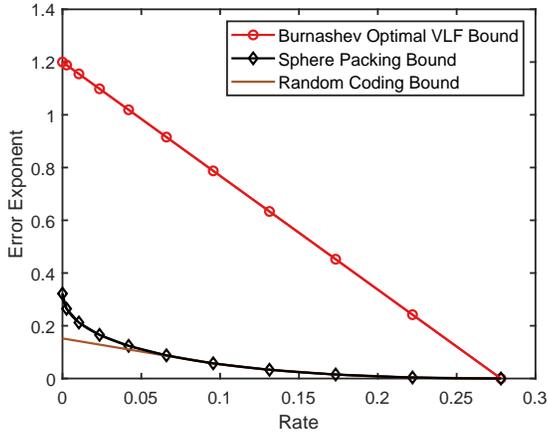}
    \caption{Figure shows the optimal error exponents of variable-length feedback codes shown by Burnashev along with upper bound and lower bound on fixed-length feedback codes i.e., sphere packing bound and random coding bound for a BSC with cross-over probability $p = 0.2$.}
    \label{fig:classical_channel_code}
\end{figure}

\begin{example} 
\label{ex:aflf_code}
Consider a Binary Symmetric Channel (BSC) with cross-over probability $p = 0.2$. Fig.~\ref{fig:classical_channel_code} shows Burnashev's optimal reliability function, the random coding bound which is a lower bound for fixed-length codes and the sphere packing bound which is an upper bound for fixed-length codes. We can see that the variable-length codes significantly improve the error exponents achieved by the class of fixed-length codes even in the presence of feedback. We shall return to this example to illustrate how one can go from the fixed-length error exponent curve to Burnashev's optimal error exponent curve in an almost-fixed-length manner.
\end{example}


Another problem at the core of information theory and statistical learning is hypothesis testing. Specifically, we consider binary hypothesis testing where a decision-maker has to choose between two hypotheses, which correspond to two possible distributions of the observed samples. There is a large body of literature on the asymptotic analysis of type-I and type-II errors as the (expected) number of samples $n$ grows. In the fixed-length regime, the number of samples is almost-surely bounded by $n$, and the error exponents of the two types of errors can only be traded-off against each other. The sequential hypothesis tests can achieve both exponents simultaneously. The sequential hypothesis test resolves the trade-off between error-types by allowing the number of samples collected to be a random number with bounded expected value. The error exponents in both variants of hypothesis testing are well-known and understood~\cite{error_exp_blahut, Error_exp_tuncel, Wald48, Chernoff59, CoverBook2nd}. The second part of the paper shows that the optimal type-I and type-II error exponents can be achieved with hypothesis tests whose stopping time is kept fixed for almost all sample paths except for an exponentially rare set. The following example shows a significant gap in the performance of the two settings of hypothesis tests.

\begin{figure}[!htb]
    \includegraphics[width=0.48\textwidth]{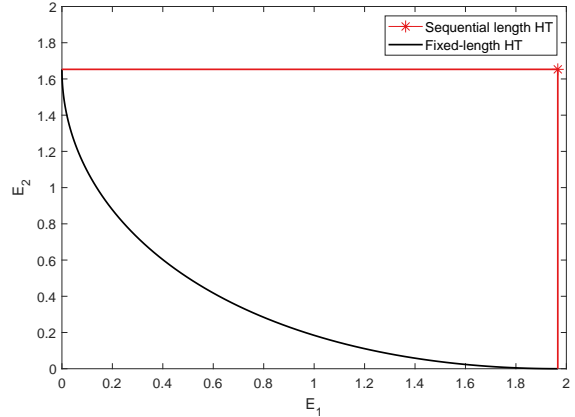}
    \caption{Figure shows the optimal error exponents of fixed-length hypothesis test and sequential hypothesis test for Bernoulli samples with parameters given by $p_1 = 0.9$ under $H_1$ and $p_2 = 0.2$ under $H_2$.}
    \label{fig:classical_HT}
\end{figure}

\begin{example} 
\label{ex:al_ht}
Consider $H_1: X \sim \ber(0.9)$ and $H_2: X \sim \ber(0.2)$. Fig.~\ref{fig:classical_HT} shows the optimal error exponents in both fixed-length and sequential setting. We can see that the sequential hypothesis test provides a significant improvement over the fixed-length hypothesis testing. We shall return to this example to illustrate how one can go from the fixed-length curve to the sequential curve.
\end{example}

To summarize our contributions are as follows:
\begin{itemize}
    \item 
    We introduce a new class of $(\gamma, K)$ almost-fixed-length channel codes with feedback for $\gamma \in [0,1]$ and $K \in \mbb{N}$. We show that the performances achieved by the two settings, namely variable-length setting and fixed-length setting are the extremities of a continuum of performance curves achieved by almost-fixed-length coding strategies as we vary $\gamma$ and $K$.
    
    \item
    We provide a lower bound to the maximum achievable error exponent by almost-fixed-length channel codes. To obtain the lower bound, we consider a two-phase $(\gamma, K)$-almost-fixed-length channel codes with feedback based where the first phase is a truncated Yamamoto-Itoh strategy~\cite{YI_1979} and the second phase is a random code. We show provide converse to almost-fixed-length feedback codes which relates to the converse of the error-erasure feedback codes with feedback.
    
    \item
    We show that Burnashev's optimal error exponent can be achieved with $(\gamma, K)$ almost-fixed-length channel codes when $\gamma = 0$ and when $K$ exceeds our lower bound implying that the flexibility needed in stopping time need not be significant.
    
    \item
    We introduce a new class of $(\gamma, K)$ hypothesis tests for $\gamma \in [0,1]$ and $K \in \mbb{N}$. We provided matching upper and lower bound for the region of all feasible error exponents for the class of $(\gamma, K)$ hypothesis tests.
\end{itemize}

\subsection{Related Work}

Channel codes with probabilistic delay constraints were considered by Altug et~al.~in~\cite{yucel_allerton_2015}, where they show that if the constraint on expected stopping time  is replaced by a probabilistic one then the first order gain in rate ceases to exist. However, these works fail to notice the improvement in the error exponent achieved by codes under probabilistic delay constraints such as the class of almost-fixed-length  codes over the class of fixed-length codes. Our results show that it is possible to achieve Burnashev's optimal error exponent with finite and bounded block-length as $\gamma$ approaches zero. As shown in Fig.~\ref{fig:classical_channel_code}, almost-fixed-length codes indeed bridge a significant gap between in an almost-fixed-length manner.

For achievability, we propose a simple construction for a two-phase almost-fixed-length feedback channel code which builds upon fixed-length feedback codes with error-erasure decoder considered by Forney in~\cite{Forney1968}, Telatar and Gallger in~\cite{erasure_exp_telatar1994} and more recently by Nakiboglu and Zheng in~\cite{Nakiboglu_error_erasure_feedback}. In the first phase we utilize an error-erasure code and whenever an erasure is declared we proceed to the second phase which consists of a fixed-length channel code. Leveraging the bounds on error-erasure exponents, we provide upper and lower bounds on the optimal error exponents achievable in an almost-fixed-length manner.

We propose a simple two-phase hypothesis test using which the overall reliability is increased significantly and the trade-off between type-I and type-II error exponents is relaxed. Our converse proof closely follows a pair of papers by Grigoryan et. al.~\cite{multi_hyp_rej} and Sason~\cite{mod_dev_HT_Sason} on hypothesis testing with rejection. Finally, we improve upon our work in~\cite{7541591} in the following ways. We extend the notion of almost-fixed-length tests to channel codes with upper and lower bounds on optimal error exponents. We also provide a matching lower bound for the error exponents achieved by almost-fixed-length hypothesis tests.

\underline{Notation:} For a set $S$ and scalar $a \in \mathbb{R}$, $a+S$ denotes the set $\{x + a: x \in S \}$ and  $aS$ denotes the set $\{ax: x \in S \}$. For sets $S_1, S_2$, $S_1 \times S_2$ denotes the set $\{(x_1,x_2):x_1 \in S_1, x_2 \in S_2\}$. Finally, the Kullback--Leibler (KL) divergence between two probability density functions $\P_1(\cdot)$ and $\P_2(\cdot)$ on space $\mathcal{X}$ is defined as $\kl{\P_1}{\P_2}=\sum_{\mathcal{X}} \P_1(x) \log\frac{\P_1(x)}{\P_2(x)}$,
with the convention $0 \log \frac{a}{0}=0$ and $b \log \frac{b}{0}=\infty$ for $a,b\in [0,1]$ with $b\neq 0$.

Rest of the paper is organized as follows. We introduce three regimes of channel codes including fixed-length feedback codes, variable-length feedback codes and almost-fixed-length feedback codes in Section~\ref{sec:channel_codes}. We provide an upper and lower bound to the maximum achievable error exponent by the class of almost-fixed-length strategy in Section~\ref{sec:aflf_code}. Similar to channel codes, we introduce three regimes of hypothesis in Section~\ref{sec:HT} and provide matching upper and lower bounds for the achievable errors for the class of almost-fixed-length hypothesis tests. We conclude the paper in Section~\ref{sec:conclusion}

\section{Types of Channel Codes}

\label{sec:channel_codes}

Consider a discrete memoryless channel (DMC) with input alphabet $\mc{X}$, output alphabet $\mc{Y}$ and a sequence of conditional output distributions $\{\P_{Y_n|X_1^n Y_1^{n-1}}\}_{n \geq 1}^{\infty}$ which satisfy the following
\begin{align}
\P_{Y_n|X_1^n Y_1^{n-1}}(y_n|x_1^n y_1^{n-1})
=
\P_{Y|X}(y_n|x_n) \quad \forall n \in \mbb{N}.
\end{align}
We assume that the feedback channel is of infinite capacity, noiseless, and delay-free, i.e., the input of the feedback channel is observed at the transmitter before transmission of $X_n$ at each time $n \in \mbb{N}$.

We discuss three classes of channel codes with feedback in the following sections: fixed-length feedback channel code, variable-length feedback channel code, and almost-fixed-length feedback channel code. These channel codes differ over the amount of variability in their stopping time. At one extreme, we have fixed-length feedback channel codes whose stopping times are bounded almost-surely. At the other extreme, we have variable-length feedback channel codes whose stopping times are bounded only in expectation. We define a new class of channel codes called almost-fixed-length feedback channel code whose stopping times are bounded with a high probability. The stopping time exceeds this upper bound with an exponentially small probability. 

\subsection{Fixed-Length Feedback Channel Codes}

\begin{definition}
\label{def:flf_code}
An $(\ell,M,\epsilon)$ fixed-length feedback (FLF) code, where $\ell, M \in \mbb{N}$, and $\epsilon \in (0,1)$,  is defined by:
\begin{enumerate}[label=(\roman*)]
\item
A common randomness $U \in \mc{U}$, with a probability distribution $\P_U$, whose realization is used to initialize the encoder and the decoder before the start of transmission.

\item
A sequence of encoders $f_n : \mathcal{U} \times \{1,\ldots, M \} \times \mathcal{Y}^{n-1}  \rightarrow \mathcal{X}$ for $n \in \mbb{N}$ defining the channel inputs
\begin{align}
X_n = f_n(U, W, Y^{n-1}),
\end{align}
where $W \in \{1,\ldots, M\}$ is the equiprobable message.

\item
A sequence of decoders $g_n : \mathcal{U} \times \mathcal{Y}^n \rightarrow \{1,\ldots, M \}$ for $n \in \mbb{N}$ providing an estimate of $W$ at each time $n$.

\item
A stopping time $\tau \in \mbb{N}$ which  satisfies
\begin{align}
\label{eq:stopping_time_flf}
\tau &= \ell \quad \text{a.s.}
\end{align}
The final estimate is computed at time $\tau$ and it is given by
\begin{align}
\hat{W}_{\tau} : = g_{\tau}(U, Y^{\tau})
\end{align}
such that the error probability satisfies
\begin{align}
\P(\hat{W}_{\tau} \neq W) \leq \epsilon.
\end{align}
\end{enumerate}

\end{definition}

\begin{definition}
A rate-reliability pair $(R,E)$ is said to be achievable in \textit{fixed-length manner with feedback} if for any $\delta > 0$ there exists an $\ell(\delta) \in \mbb{N}$ such that for all $\ell \geq \ell(\delta)$ there is an $(\ell, M_{\ell}, \epsilon_{\ell})$ FLF code which satisfies
\begin{align}
&M_{\ell} \geq 2^{\ell R},\\
&\epsilon_{\ell} \leq 2^{-\ell(E-\delta)}.
\end{align} 
For a given rate $R$ below capacity, the reliability function $\mc{E}_{\text{FLF}}(R) $ is defined as the best achievable error exponent at rate $R$ in a fixed-length manner with feedback.
\end{definition}

The following fact characterizes an upper bound and lower bound on the optimal achievable reliability in a fixed-length manner with feedback.

\begin{fact}
\label{fact:flf_bounds} 
For FLF codes, the optimal achievable reliability as a function of rate $R$ can bounded above and below as follows
\begin{align}
E_{\text{r}}(R) \leq \mc{E}_{\text{FLF}}(R) \leq E_{\text{H}}(R),
\end{align}
where $E_{\text{H}}(R)$ is an upper established by Haroutunian in~\cite{haroutunian_1977} and $E_r(R)$ is the random coding exponent demonstrated in~\cite{Gallager_book}. The Haroutunian exponent $E_{\text{H}}(R)$ is the best known upper bound to the error exponent for fixed-length coding with feedback. The Haroutunian exponent $E_{\text{H}}(R)$ is strictly larger than the sphere packing exponent $E_{\text{sp}}(R)$ (demonstrated in~\cite{Gallager_book}) for the class of non-symmetric channels such as the Z-channel and coincides with $E_{\text{sp}}(R)$ for the class of symmetric channels such as the Binary Symmetric Channel (BSC) and the Binary Erasure Channel (BEC)~\cite{upper_bound_nakiboglu}.  
\end{fact}

\subsection{Variable-Length Feedback Channel Codes}

\begin{definition}
\label{def:vlf_code}
An $(\ell,M,\epsilon)$ variable-length feedback (VLF) code, where $\ell, M \in \mbb{N}$, and $\epsilon \in (0,1)$,  is defined similarly to FLF codes with an exception that condition (iv) in Definition~\ref{def:flf_code} is replaced by:

\noindent
(iv)$^{\prime}$ A random stopping time $\tau \in \mbb{N}$ which satisfies
\begin{align}
\label{eq:stopping_time_vlf}
\expe[\tau] &\leq \ell \quad \text{a.s.}
\end{align}
\begin{addmargin}[8mm]{1mm}
The final estimate is computed at time $\tau$ and it is given by
\begin{align}
\hat{W}_{\tau} : = g_{\tau}(U, Y^{\tau})
\end{align}
such that the error probability satisfies
\begin{align}
\P(\hat{W}_{\tau} \neq W) \leq \epsilon.
\end{align}
\end{addmargin}

%
%
%
%
%

\end{definition}

\begin{definition}
A rate-reliability pair $(R,E)$ is said to be achievable in \textit{variable-length manner with feedback} if for any $\delta > 0$ there exists an $\ell(\delta) \in \mbb{N}$ such that for all $\ell \geq \ell(\delta)$ there is an $(\ell, M_{\ell}, \epsilon_{\ell})$ VLF code which satisfies
\begin{align}
&M_{\ell} \geq 2^{\ell R},\\
&\epsilon_{\ell} \leq 2^{-\ell(E-\delta)}.
\end{align} 
For a given rate $R$ below capacity, the reliability function $\mc{E}_{\text{VLF}}(R) $ is defined as the best achievable error exponent at rate $R$ in a variable-length manner with feedback.
\end{definition}

The following fact characterizes the optimal achievable reliability in a variable-length manner with feedback.

\begin{fact}
\label{fact:burnashev_vlf}
Burnashev in~\cite{Burnashev76} established that for the class of VLF codes, the optimal achievable reliability as a function of rate $R$ is given by
\begin{align}
\mc{E}_{\text{VLF}}(R) = C_1\left( 1 -\frac{R}{C} \right),
\end{align}
where $C$ denotes the capacity of the DMC and $C_1$ denotes the maximal KL-divergence between the conditional output distributions given any two inputs, i.e., 
\begin{align}
C_1 := \max_{x, x^{\prime} \in \mc{X}} D(\P_{Y|X}(\cdot|x)|| \P_{Y|X}(\cdot|x^{\prime})).
\end{align}
For BSC, Burnashev's error exponent is strictly greater than the sphere packing exponent.
\end{fact}

In summary, an optimal FLF code can at most achieve only the Haroutunian exponent $E_H(R)$ or sphere packing exponent $E_{\text{sp}}(R)$ for symmetric channels. In contrast, a VLF code can achieve the Burnashev's optimal error exponent which is significantly larger than the upper bounds for error exponent of FLF code. Example~\ref{ex:aflf_code} along with Fig.~\ref{fig:classical_channel_code} illustrates this.

\subsection{Almost-fixed-length Feedback Channel Codes}

We introduce a new class of channel codes for which the maximum length of the code is bounded, but the stopping time has some variability. By construction, the stopping time of an $(\ell,M,\gamma, K, \epsilon)$ almost-fixed-length channel codes exceeds $\ell$ with exponentially small probability, and on sample paths where the code length exceeds $\ell$, the maximum length of the code is almost-surely bounded by $K\ell$.

\begin{definition}
An $(\ell,M,\gamma, K, \epsilon)$ almost-fixed-length feedback (AFLF) code, where $\ell, M, K \in \mbb{N}$, $\gamma \geq 0$, and $\epsilon \in (0,1)$, is defined similarly to FLF codes with an exception that condition (iv) in Definition~\ref{def:flf_code} is replaced by:

\noindent
(iv)$^{\prime\prime}$ A random stopping time $\tau \in \mbb{N}$ which satisfies
\begin{align}
\label{eq:stopping_time_cc1}
\P(\tau > \ell) &\leq 2^{-\gamma \ell},
\\
\label{eq:stopping_time_cc2}
\tau &\leq K\ell \quad \text{a.s.}
\end{align}
\begin{addmargin}[8mm]{1mm}
The final estimate is computed at time $\tau$ is given by
\begin{align}
\hat{W}_{\tau} : = g_{\tau}(U, Y^{\tau})
\end{align}
such that the error probability satisfies
\begin{align}
\P(\hat{W}_{\tau} \neq W) \leq \epsilon.
\end{align}
\end{addmargin}

%
%
%
%
%

\end{definition}

\begin{remarks}
The support of the stopping time $\tau$ for any $(\ell, M, \gamma, K, \epsilon)$ AFLF channel code is $\{\ell, K\ell\}$. The $k$-th moment of the stopping time for an $(\ell, M, \gamma, K, \epsilon)$ AFLF channel code is finite for any $k > 1$, and it can be upper and lower bounded as 
\begin{align}
    &\ell^k(1-2^{-\gamma \ell}) \leq \expe[\tau^k] \leq \ell^k + K^k\ell^k 2^{-\gamma \ell},
\end{align}
Hence, we have 
\begin{align}
    \lim_{\ell \to \infty}\frac{\expe[\tau^k]}{\ell^k} = 1.
\end{align}
This implies the stopping time concentrates around $\ell$ with an exponentially decaying tail and the variability of the stopping time for AFLF codes decreases as $\ell$ increases. 
\end{remarks}

\begin{remarks}
\label{remm:special_case_aflf}
From the definition of AFLF code we make the following observations:
\begin{enumerate}[label=(\roman*)]

\item
The FLF codes are a special case of AFLF codes where $\gamma = \infty$ and $K = 1$ and hence $\P(\tau > \ell) = 0$.

\item
AFLF codes are a special case of VLF codes, where the condition on the stopping time
\begin{align}
\expe[\tau] \leq \ell,
\end{align}
is replaced by more stringent conditions given in equations~\eqref{eq:stopping_time_cc1} and~\eqref{eq:stopping_time_cc2}. In other words, AFLF codes not only require the stopping time $\tau$ to be bounded in expectation but also require the probability that $\tau$ exceeds $\ell$ to be exponentially small.

\item
AFLF codes are a special case of VLF$^{\ast}$ codes considered by Altug et~al.~in~\cite{yucel_allerton_2015}, where the condition on the stopping time 
\begin{align}
\label{eq:yucel_condition}
\min\{n\in \mbb{N}: \P(\tau > n) \leq \epsilon_d\} \leq \ell,
\end{align}
for some $\epsilon_d \in (0,1)$ is replaced by more stringent conditions given in equations~\eqref{eq:stopping_time_cc1} and~\eqref{eq:stopping_time_cc2}. In other words, AFLF codes not only require the probability that $\tau$ exceeds $\ell$ to be less than a fixed threshold $\epsilon_d$ but also require the threshold to decay exponentially in $\ell$.

\end{enumerate}

\end{remarks}

\begin{definition}
A rate-reliability pair $(R,E)$ is said to be achievable in a $(\gamma, K)$-\textit{almost-fixed-length manner with feedback} if for any $\delta > 0$ there exists an $\ell(\delta) \in \mbb{N}$ such that for all $\ell \geq \ell(\delta)$ there is a $(\ell, M_{\ell}, \gamma, K, \epsilon_{\ell})$ AFLF code which satisfies
\begin{align}
&M_{\ell} \geq 2^{\ell R},\\
&\epsilon_{\ell} \leq 2^{-\ell(E-\delta)}.
\end{align} 
For a given rate $R$ below capacity, the reliability function $\mc{E}_{\text{AFLF}}(R, \gamma, K) $ is defined as the best achievable error exponent at rate $R$ in a $(\gamma, K)$-almost-fixed-length manner with feedback. 
\end{definition}


The following corollary characterizes an upper bound and lower bound on the optimal achievable reliability in an almost-fixed-length manner with feedback.

\begin{corollary}
\label{coro:aflf_basic_bounds}
For the class of AFLF codes, for all $0 \leq \gamma < \infty$ and $K \in \mbb{N}$, the optimal achievable reliability as a function of rate $R$  can be bounded as follows
\begin{align}
    E_{\text{r}}(R) \leq \mc{E}_{\text{AFLF}}(R, \gamma, K) \leq C_1\left( 1 -\frac{R}{C} \right),
\end{align}
and for $\gamma = \infty$ and $K \in \mbb{N}$, we have
\begin{align}
   E_{\text{r}}(R) \leq \mc{E}_{\text{AFLF}}(R, \infty, 1) \leq E_{\text{H}}(R). 
\end{align}
\end{corollary}

Remark~\ref{remm:special_case_aflf}~(i) and~(ii) imply 
\begin{align}
\mc{E}_{\text{FLF}}(R) = \mc{E}_{\text{AFLF}}(R, \infty, 1) \leq \mc{E}_{\text{AFLF}}(R, \gamma, K), \\
\mc{E}_{\text{AFLF}}(R, \gamma, K) \leq \mc{E}_{\text{AFLF}}(R, 0, \infty) = \mc{E}_{\text{VLF}}(R).
\end{align}
Combining Fact~\ref{fact:flf_bounds} and  Fact~\ref{fact:burnashev_vlf} with the above equations we obtain Corollary~\ref{coro:aflf_basic_bounds}.

\section{Rate-Reliability of Almost-Fixed-Length Feedback Codes}

\label{sec:aflf_code}

Next we provide our main results for almost-fixed-length feedback codes.

\subsection{Achievability: Construction of Almost-Fixed-Length Feedback Codes}

In this section we show that AFLF codes can be easily constructed from existing channel coding strategies such as error-erasure channel codes. More specifically, we utilize error-erasure codes proposed by Forney in~\cite{Forney1968}, Telatar and Gallger in~\cite{erasure_exp_telatar1994} and more recently by Nakiboglu and Zheng in~\cite{Nakiboglu_error_erasure_feedback}. Next we formally define error-erasure codes. 

\begin{definition}
\label{def:error_erasure_code}
An $(\ell,M,\epsilon, \epsilon_{\mbf{x}})$ fixed-length error-erasure feedback code, where $\ell, M \in \mbb{N}$, and $\epsilon, \epsilon_{\mbf{x}} \in (0,1)$,  is defined by:
\begin{enumerate}[label=(\roman*)]
\item
A common randomness $U \in \mc{U}$, with a probability distribution $\P_U$, whose realization is used to initialize the encoder and the decoder before the start of transmission.

\item
A sequence of encoders $f_n : \mathcal{U} \times \{1,\ldots, M \} \times \mathcal{Y}^{n-1}  \rightarrow \mathcal{X}$ for $n \in \mbb{N}$ defining the channel inputs
\begin{align}
X_n = f_n(U, W, Y^{n-1}),
\end{align}
where $W \in \{1,\ldots, M\}$ is the equiprobable message.

\item
A sequence of decoders $g_n : \mathcal{U} \times \mathcal{Y}^n \rightarrow \{1,\ldots, M \}\cup \{e\}$ for $n \in \mbb{N}$ providing the best estimate of $W$ in $\{1,\ldots,M\}$ or declare an erasure $e$ at time $n$.

\item
A stopping time $\tau \in \mbb{N}$ which  satisfies
\begin{align}
\label{eq:stopping_time_ee}
\tau &= \ell \quad \text{a.s.}
\end{align}
The final estimate is computed at time $\tau$ and it is given by
\begin{align}
\hat{W}_{\tau} : = g_{\tau}(U, Y^{\tau})
\end{align}
such that the error probability satisfies
\begin{align}
\P(\hat{W}_{\tau} \neq W) - \P(\hat{W}_{\tau} = e) \leq \epsilon,
\end{align}
and 
the erasure probability satisfies
\begin{align}
\P(\hat{W}_{\tau} = e) \leq \epsilon_{\mbf{x}}.
\end{align}
\end{enumerate}

\end{definition}

\begin{definition}
A rate-reliability pair $(R,E, E_{\mbf{x}})$ is said to be achievable in \textit{fixed-length manner with feedback and error-erasure decoding} if for any $\delta > 0$ there exists an $\ell(\delta) \in \mbb{N}$ such that for all $\ell \geq \ell(\delta)$ there is a $(\ell, M_{\ell}, \epsilon_{\ell}, \epsilon_{\ell, \mbf{x}})$ fixed-length error-erasure feedback code which satisfies
\begin{align}
&M_{\ell} \geq 2^{\ell R},\\
&\epsilon_{\ell} \leq 2^{-\ell(E-\delta)},\\
&\epsilon_{\ell, \mbf{x}} \leq 2^{-\ell(E_{\mbf{x}}-\delta)}.
\end{align} 
For a given rate $R$ below capacity and $E_{\mbf{x}} \geq 0$, the error reliability function $\mc{E}_{\text{ee}}(R, E_{\mbf{x}}) $ is defined as the best achievable error exponent at rate $R$ and erasure exponent $E_{\mbf{x}}$ in a fixed-length manner with feedback and error-erasure decoding.
\end{definition}

Next, we construct AFLF codes using the class of error-erasure codes and FLF codes.

\begin{proposition}
\label{prop:basic_two_phase_strategy}
Consider a $(\ell, M, \epsilon_1, \epsilon_{\mbf{x}})$ fixed-length error-erasure feedback code and a $((K-1)\ell, M, \epsilon_2)$ fixed-length feedback code, applied sequentially in a two phase strategy as given below:
\begin{enumerate}[label=(\roman*)]
\item
\underline{Phase I:} For all time instants $n \leq \ell$, the message $W$ is transmitting using the encoding functions of $(\ell, M, \epsilon_1, \epsilon_{\mbf{x}})$ fixed-length error-erasure feedback code. At $n = \ell$, use the decoding function of the $(\ell, M, \epsilon_1, \epsilon_{\mbf{x}})$ code to obtain an estimate $\hat{W}_{\ell}$ of the message $W$. If $\hat{W}_{\ell} \neq e$ then stop and if $\hat{W}_{\ell} = e$, then proceed to Phase II.

\item
\underline{Phase II:} For all time instants $\ell < n \leq K\ell$, discard the previous channel observations and transmit the message $W$ using the encoding functions of $((K-1)\ell, M, \epsilon_2)$ FLF code. At $n = K\ell$, use the decoding function of the $((K-1)\ell, M, \epsilon_2)$ FLF code to obtain an estimate $\hat{W}_{K\ell}$ of the message $W$. 
\end{enumerate}
The resulting two-phase strategy is an $(\ell, M)$ almost-fixed-length feedback code which satisfies
\begin{align}
\P(\tau > \ell) &\leq \epsilon_{\mbf{x}},
\\
\tau &\leq K\ell \quad \text{a.s.}
\\
\P(\hat{W}_{\tau} \neq W) &\leq \epsilon_1 + \epsilon_2.
\end{align}
\end{proposition}



We use Proposition~\ref{prop:basic_two_phase_strategy} to construct AFLF channel codes. Next two theorems provide a lower bound to the optimal achievable reliability in an almost-fixed-length manner with feedback when $\gamma = 0$ and when $\gamma > 0$.

\begin{theorem}[Special Case: $\gamma = 0$]
\label{thm:achv_aflf_gamma_0}
For the class of AFLF codes, for $\gamma =0$ and $K \in \mbb{N}$, the  optimal achievable reliability as a function of rate $R$ can be lower bounded as
\begin{align}
&\mc{E}_{\text{AFLF}}(R, 0, K) 
\\
&\geq
\min \left\{ C_1 \left( 1 - \frac{R}{C}\right), 
(K-1) E_{\text{r}} \left(\frac{R}{(K-1)}\right)
\right \}.
\end{align}
Furthermore, for all $K \geq K^{\ast} := 1+\frac{C_1}{E_\text{r}(0)}$ we have 
\begin{align}
\mc{E}_{\text{AFLF}}(R, 0, K) = C_1 \left( 1 - \frac{R}{C}\right).
\end{align}
\end{theorem}
Proof of the above theorem is provided in Appendix~\ref{proof:achv_aflf_gamma_0}. Proof of the theorem is obtained by constructing a two-phase strategy where the first phase is a truncated Yamamoto-Itoh strategy~\cite{YI_1979} and the second phase is a random code. Theorem~\ref{thm:achv_aflf_gamma_0} shows that the optimal Burnashev bound $\mc{E}_{\text{VLF}}(R)$ can be achieved for any rate $R$ below capacity using a code with bounded and finite maximum code length and limited variability in the stopping time. The class of VLF codes also achieves the optimal Burnashev bound; however, these codes' stopping time is bounded only in expectation.  Unlike AFLF codes,  VLF codes may occasionally have a very large stopping time.

\begin{theorem}[Strictly Positive $\gamma > 0$]
\label{thm:achv_aflf_any_gamma}
For any rate $R$ below capacity, let
\begin{align}
\alpha^{\ast}(R, \gamma) := \frac{R}{g^{-1}(\frac{\gamma}{R})},
\end{align}
where $g(a) = \frac{E_{\text{r}}(a)}{a}$. Let $x, x'$ denote two inputs which maximize the KL-divergence between the conditional output distributions given the channel input. For the class of AFLF codes, for $0 < \gamma < E_{\text{r}}(R)$ and $K \in \mbb{N}$, the  optimal achievable reliability as a function of rate $R$ can be lower bounded as
\begin{align}
\label{eq:aflf_lower_bound}
\mc{E}_{\text{AFLF}}(R, \gamma, K) 
\geq
\min \left\{
E_{\text{ee}}^{\prime}(R, \gamma), (K-1)E_{\text{r}}\left( \frac{R}{K-1}\right)
\right\} ,
\end{align}
where define
\begin{align}
&E_{\text{ee}}^{\prime}(R, \gamma) 
: = 
\\
&\max_{\alpha \in [\alpha^{\ast}(R, \gamma), 1]} 
\max_{\substack{
\lambda \in [0,1] 
\\  \left( 1 - \alpha \right) D(\P^{(\lambda)}||\P_{Y|X}(\cdot|x))
\geq \gamma}}E_{\text{ee}}^{\prime\prime}(\alpha, \lambda, R),
\end{align}
and define
\begin{align}
E_{\text{ee}}^{\prime\prime}(\alpha, \lambda, R)
: =  \alpha   E_{\text{r}}\left(\frac{R}{\alpha}\right)+ \left( 1 - \alpha \right) D(\P^{(\lambda)}||\P_{Y|X}(\cdot|x^{\prime})),
\end{align}
where for any $\lambda \in [0, 1]$, the $\lambda$-tilted distribution $\P^{(\lambda)}$ is given by
\begin{align}
\P^{(\lambda)}(y) 
:=
\frac{\P_{Y|X}^{1-\lambda}(y|x) \P_{Y|X}^{\lambda}(y|x^{\prime})}{ \sum_{a \in \mathcal{Y}} \P_{Y|X}^{1-\lambda}(a|x) \P_{Y|X}^{\lambda}(a|x^{\prime})}, \quad \forall \, y \in \mathcal{Y}.
\end{align}
Additionally, for $\gamma > E_{\text{r}}(R)$ the lower bound on optimal achievable reliability as a function of rate $R$ given by equation~\eqref{eq:aflf_lower_bound} reduces to $\mc{E}_{\text{AFLF}}(R, \gamma , K) \geq E_{\text{r}}(R)$.
\end{theorem}
Proof of the above theorem is provided in
Appendix~\ref{proof:achv_aflf_any_gamma}. Proof of the theorem is obtained by constructing a two-phase strategy where the first phase is a truncated Yamamoto-Itoh strategy~\cite{YI_1979} and the second phase a random code.

\begin{remarks}
Polyanskiy et~al.~in~\cite{yuri_feedback_2011} show that for VLF channel codes feedback improves first and second-order terms of the logarithm of the maximum number of messages that can be transmitted with a non-vanishing error probability. Altug et~al.~in~\cite{yucel_allerton_2015} show that channel codes with more stringent constraints on stopping time, designed to reduce its variability, provide no such improvement even in the presence of feedback. Specifically, if the constraint on the expected value of the stopping time of VLF codes is replaced by a probabilistic one, given by equation~\eqref{eq:yucel_condition}, the first order gain in rate ceases to exist. However, these works fail to notice the improvement in the error exponent achieved by codes under probabilistic delay constraints such as AFLF codes over FLF codes.  Furthermore, Theorem~\ref{thm:achv_aflf_gamma_0} and \ref{thm:achv_aflf_any_gamma} show that the class of AFLF codes can improve the error exponent significantly and that it is possible to achieve Burnashev's optimal error exponent with finite and bounded code length as $\gamma$ approaches zero.
\end{remarks}

\begin{remarks}
Gopala et~al.~in~\cite{arq_erasure_gopala2007} propose IR-ARQ strategy  over Automatic ReQuest (ARQ) memoryless channels when the maximum number of ARQ rounds $K$ is constrained. At the end of each ARQ round, the decoder either decodes the transmitted message or declares an erasure. In case of an erasure, the transmitter proceeds to the next ARQ round. In the final round, the decoder employs the maximum likelihood decoding rule to decide on the transmitted message. Also, the decoder does not discard the received channel output sequences in the case of an erasure and uses the received sequences of all the ARQ rounds jointly to decode the transmitted message. The error exponent $E_{\text{IR}}(R,K)$ achieved by IR-ARQ strategy (Theorem~3 in~\cite{arq_erasure_gopala2007}) satisfies
\begin{align}
E_{\text{IR}}(R,K) \geq \min \left\{ E_f(R), KE_r\left( \frac{R}{K} \right) \right\},
\end{align}
where exponent $E_f(R)$ denotes Forney's decision feedback exponent which is defined as the largest possible error exponent for a error-erasure feedback code while the erasure exponent is positive~\cite{Forney1968}. For a BSC, if $K \geq \frac{E_f(0)}{E_r(0)} $, where $E_r(R)$ is the BSC random coding exponent (Lemma~4 in~\cite{arq_erasure_gopala2007}), which implies $E_{\text{IR}}(R,K) \geq E_f(R)$, for all $0 \leq R \leq C$. Hence, IR-ARQ achieves the Forney's decision feedback exponent for $K \geq \frac{E_f(0)}{E_r(0)}$, while AFLF codes with $\gamma = 0$ achieves Burnashev's optimal VLF exponent for $K \geq 1+\frac{C_1}{E_r(0)}$.

Fig.~\ref{fig:forney_bound} plots Burnashev's optimal VLF exponent with Forney's decision feedback bound, sphere packing bound and random coding bound for a BSC with crossover probability $p = 0.2$. The figure shows that Burnashev's optimal VLF bound is greater than Forney's decision feedback bound. This also implies that the joint decoding of channel outputs utilized by the IR-ARQ strategy reduces the code's length in the worst case, but it comes at the cost of a smaller error exponent guarantee. 

\begin{figure}[!htb]
    \includegraphics[width=0.45\textwidth]{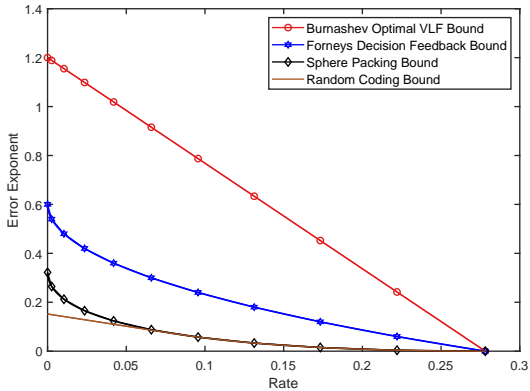}
    \caption{Figure shows Forney's decision feedback bound and Burnashev's optimal VLF exponent with the upper bound and the lower bound on FLF codes, i.e., sphere packing bound and random coding bound for a BSC with cross-over probability $p = 0.2$.}
\label{fig:forney_bound}
\end{figure}
\end{remarks}

\begin{figure}[!htb]
    \includegraphics[width=0.45\textwidth]{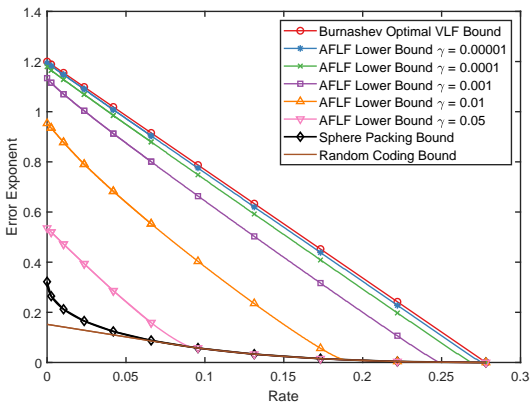}
    \caption{Figure shows the error exponents achieved by AFLF codes when $K \geq K^{\ast}=7$ for values of $\gamma$ approaching zero. The AFLF bounds are compared with the Burnashev's optimal VLF exponent along with upper bound and lower bound on FLF codes i.e., sphere packing bound and random coding bound for a BSC with cross-over probability $p = 0.2$. }
\label{fig:aflf_channel_code}
\end{figure}

\noindent
\textbf{Example 1 (Revisited).}
For the setup considered in Example~\ref{ex:aflf_code},  Fig.~\ref{fig:aflf_channel_code} shows the lower bound for the optimal AFLF error exponent $\mc{E}_{\text{AFLF}}(R, \gamma, K)$ obtained in equation~\eqref{eq:aflf_lower_bound} of Theorem~\ref{thm:achv_aflf_any_gamma}. We see that as $\gamma$ decreases, the rate-reliability curve improves. In particular, for any rate $R$ below capacity it is possible to achieve an error exponent arbitrarily close to Burnashev's optimal VLF exponent $E_{\text{VLF}}(R)$ in an almost-fixed-length manner by selecting $\gamma$ close to zero. 
Note that for any $R$ for which $E_{\text{r}}(R) < \gamma$ the lower bound for $\mc{E}_{\text{AFLF}}(R, \gamma, K)$ coincides with $E_{\text{r}}(R)$ as in Fig.~\ref{fig:aflf_channel_code}.

\subsection{Converse for Almost-Fixed-Length Feedback Codes}

Our converse upper bounds the optimal error exponent of a $(\ell, M, \gamma, K, \epsilon)$ AFLF code with that of a fixed-length error-erasure feedback code where the probability of erasure approaches zero exponentially fast with an exponent at most $\gamma$. More specifically, given an $(\ell, M, \gamma, K, \epsilon)$ AFLF code we can construct a fixed-length error-erasure code by declaring an erasure whenever after $\ell$ channel uses $\tau > \ell$. Such a code is a $(\ell, M, 2^{-\gamma \ell}, \epsilon)$ fixed-length feedback code with error-erasure decoder. As a consequence we obtain the following result.

\begin{proposition}[Converse]
\label{prop:converse}
For the class of AFLF codes, for $\gamma \geq 0$ and $K \in \mbb{N}$, the optimal achievable reliability as a function of rate $R$ can be upper bounded as
\begin{align}
\mc{E}_{\text{AFLF}}(R, \gamma, K) \leq
\min \left\{\mc{E}_{\text{ee}}(R,\gamma), 
K E_{\text{H}} \left(\frac{R}{K}\right)
\right \},
\end{align}
where recall that $\mc{E}_{\text{ee}}(R, \gamma)$ is the optimal achievable error exponent under error-erasure decoding where the erasure exponent is $\gamma$ and $E_{\text{H}}(R)$ denotes the upper bound establised by Haroutunian in~\cite{haroutunian_1977} for the class of FLF codes.
\end{proposition}
Proof of the above proposition is provided in the Appendix~\ref{proof:converse}.

Proposition~\ref{prop:converse} implies that obtaining optimal error exponent for AFLF codes is related to the optimal error exponent $\mc{E}_{ee}(R, E_{\mbf{x}})$ for any erasure exponent $E_{\mbf{x}} > 0$. However, to obtain the optimal error exponent $\mc{E}_{ee}(R, E_{\mbf{x}})$ when erasure exponent is $E_{\mbf{x}}$ (thereby, obtain the optimal error exponent and erasure exponent trade-off), we must solve the important and challenging open problem of finding the optimal error exponent of erasure-free fixed-length feeddback codes. To this end, Nakiboglu and Zheng in~\cite{Nakiboglu_error_erasure_feedback} provide an upper bound for $\mc{E}_{ee}(R, E_{\mbf{x}})$ for any erasure exponent is $E_{\mbf{x}} >0$ and their bound is loose only as much as the best known upper bound for the error exponent of the erasure-free feedback codes is loose.

\section{Types of Hypothesis Tests}
\label{sec:HT}

In this section, we extend the notion of almost-fixed-length strategies to hypothesis testing and provide matching upper and lower bounds for error exponents achieved by binary hypothesis tests in an almost-fixed-length manner. Consider two hypotheses $H_1$ and $H_2$ which correspond to the two possible underlying distribution $\P_1$ and $\P_2$ of the samples. In other words, we have
\begin{align}
H_1 : X \sim \P_1,
\quad \text{and} \quad
H_2 : X \sim \P_2,
\end{align}
where $X$ takes values in a finite set $\mathcal{X}$. Consider collecting $\tau$ number of i.i.d samples, where $\tau$ is a random stopping time with respect to the underlying filtration given by $\sigma(X_1, \ldots, X_n)$. \add{The expectation under hypothesis $H_i$, for $i \in \{ 1, 2\}$, is denoted by $\expe_i[\cdot]$.}

A general \emph{hypothesis test} decides between $H_1$ and $H_2$, for any given $\tau$ samples by dividing the sample space $\mathcal{X}^{\tau}$ into two sets or two ``decision regions''. A decision region, denoted by $A_i^{\tau}$, is a collection of samples $X^{\tau} \in \mathcal{X}^{\tau}$ for which the test chooses $H_i$, for $i \in \{1, 2\}$. Hence, we have $A_1^{\tau} \cup A_2^{\tau} = \mc{X}^{\tau}$. The type-I error is defined as an error event that occurs when the test accepts hypothesis $H_2$ when hypothesis $H_1$ is true and its probability is given by $\P_1 \left( A_2^{\tau} \right)$. Similarly, type-II error is defined as an error event when the test accepts hypothesis $H_1$ when hypothesis $H_2$ is true and its probability is given by $\P_2 \left( A_1^{\tau} \right)$. It is known that collecting more number of samples results in an exponential reduction in these probabilities of error. This fact is characterized by two classical asymptotic results depending on the manner in which stopping time $\tau$ grows. First, we review two classical regimes of hypothesis tests based on the growth of $\tau$. Then, we introduce a new class of almost-fixed-length hypothesis tests.

\subsection{Fixed-Length Hypothesis Tests}

In this setting $\tau$ is assumed to be a bounded integer i.e., it satisfies $\tau \leq n$, where $n \in \mathbb{N}$.

\begin{definition}
The error exponents $(E_1, E_2)$ are said to be achievable in a \textit{fixed-length} manner, if for every $\delta > 0$ there exists an $N(\delta) \in \mbb{N}$ such that for all $n \geq N(\delta)$ there exists a hypothesis test satisfying the following constraints
\begin{align}
&\tau \leq n \quad \P_i-\mbox{a.s. for } i \in \{ 1, 2\},
\\
&\P_1 \left( A_2^{\tau} \right)
\leq e^{-(E_1 - \delta)n},
\\
&\P_2 \left( A_1^{\tau} \right)
\leq e^{-(E_2 - \delta)n}.
\end{align}
\end{definition}

\begin{definition} 
For any $\lambda \in [0, 1]$, the $\lambda$-tilted distribution $\P^{(\lambda)}$ with respect to $\P_1$ and $\P_2$ is given by
\begin{align}
\P^{(\lambda)}(x) 
:=
\frac{\P_1^{1-\lambda}(x) \P_2^{\lambda}(x)}{ \sum_{a \in \mathcal{X}} \P_1^{1-\lambda}(a) \P_2^{\lambda}(a)}, \quad \forall \, x \in \mathcal{X}.
\end{align}
\end{definition}

The following fact characterizes the set of all error exponents, denoted by $\mathcal{R}_{\text{FL}}$, achievable in a fixed-length manner.

\begin{fact}[Theorem~11.7.1 in \cite{CoverBook2nd}]
\label{fact:fl_ht}
The set of error exponents feasible for the class of fixed-length hypothesis tests is  given by 
\begin{align}
\mathcal{R}_{\text{FL}} = 
\left\{ (E_1, E_2): \right.
E_i \leq \kl{\P^{(\lambda)}}{\P_i}, \, i \in \{ 1, 2\},
\nonumber
\\
\hspace{1cm}
\left. \text{ for some } \lambda \in [0,1] \right\}.
\end{align}
Furthermore, the following fixed-length hypothesis test achieves the optimal error exponents on the boundary of $\mathcal{R}_{FL}$. If
\begin{align}
\begin{array}{ll}
\frac{1}{n}\sum_{i = 1}^{n} \log \frac{\P_1(X_i)}{\P_2(X_i)} \geq \alpha & \text{ stop and choose } H_1,
\\
\frac{1}{n}\sum_{i = 1}^{n} \log \frac{\P_1(X_i)}{\P_2(X_i)} < \alpha & \text{ stop and choose } H_2,
\end{array}
\end{align}
where $\alpha$ is given by
\begin{align}
\alpha = \kl{\P^{(\lambda)}}{\P_2} - \kl{\P^{(\lambda)}}{\P_1}, \, \lambda \in [0,1]. 
\end{align}
\end{fact}

\begin{definition}
Let $\lambda^{\ast}$ be such that
\begin{align}
\kl{\P^{(\lambda^{\ast})}}{\P_1} = \kl{\P^{(\lambda^{\ast})}}{\P_2}.
\end{align}
Then, the Chernoff exponent $D^{\ast}$ is defined as
\begin{align}
D^{\ast} := \kl{\P^{(\lambda^{\ast})}}{\P_1},
\end{align}
and it characterizes the optimal reliability of Bayesian hypothesis tests. In other words, $D^{\ast}$ denotes the optimal exponent that can be achieved simultaneously by both type-I and type-II errors in a fixed-length manner. 
\end{definition}

\subsection{Sequential Hypothesis Tests} 

In this setting, $\tau$ is allowed to be a random variable (potentially unbounded) such that \add{$\max \{ \expe_1[\tau], \expe_2[\tau] \} \leq n$, where $n \in \mathbb{N}$.}

\begin{definition}
The error exponents $(E_1, E_2)$ are said to be sequentially achievable, if for every $\delta > 0$ there exists an $N(\delta) \in \mbb{N}$ such that for all $n \geq N(\delta)$ there exists a hypothesis test satisfying the following constraints
\begin{align}
\label{eq:seq_1}
&\add{ \max \{\expe_1[\tau], \expe_2[\tau] \} \leq n},
\\
\label{eq:seq_2}
&\P_1 \left( A_2^{\tau} \right)
\leq e^{-(E_1 - \delta)n},
\\
\label{eq:seq_3}
&\P_2 \left( A_1^{\tau} \right)
\leq e^{-(E_2 - \delta)n}.
\end{align} 
\end{definition}

The following fact characterizes the set of all error exponents, denoted by $\mathcal{R}_{\text{seq}}$, achievable in sequential manner.

\begin{fact}[Wald and Wolfowitz~\cite{Wald48}]
The set of error exponents feasible for the class of sequential hypothesis test are given by 
\begin{align}
&\mathcal{R}_{\text{seq}} 
\nonumber
\\
= 
&\add{\{E_1: E_1 \leq \kl{\P_2}{\P_1}\} \times \{ E_2:
E_2 \leq \kl{\P_1}{\P_2} \}}.
\end{align} 
Furthermore, for $\delta >0$ the following sequential hypothesis test achieves the above error exponents arbitrarily close to optimal error exponents $(\kl{\P_2}{\P_1}, \kl{\P_1}{\P_2})$. At any instant $k \in \mathbb{N}$,
\begin{align}
\begin{array}{ll}
\sum_{i = 1}^{k} \log \frac{\P_1(X_i)}{\P_2(X_i)} \geq  \alpha & \text{stop and choose } H_1,
\\
\sum_{i = 1}^{k} \log \frac{\P_1(X_i)}{\P_2(X_i)} \leq \beta & \text{stop and choose } H_2,
\\
\beta < \sum_{i = 1}^{k} \log \frac{\P_1(X_i)}{\P_2(X_i)} < \alpha & \text{ take an extra sample}\\
 & \text{ and repeat for } k+1,
\end{array}
\end{align}
where $\alpha = (\kl{\P_1}{\P_2} - \delta)n$ and $\beta = -(\kl{\P_2}{\P_1} - \delta)n$. As $\delta$ goes to the error exponents converge to the optimal error exponents.
\end{fact}

\begin{remarks} 
Our definition of sequentially achievable error exponents, given by equations~\eqref{eq:seq_1}, \eqref{eq:seq_2}, and \eqref{eq:seq_3},  coincides with the achievable error exponents defined in~\cite{Csiszar04}. 
Alternatively, the definition can be modified such that for $n_1, n_2 \in \mbb{N}$ large enough the hypothesis test satisfies 
\begin{align}
& \expe_1[\tau] \leq n_1, \quad \expe_2[\tau] \leq n_2,
\\
&\P_1 \left( A_2^{\tau} \right)
\leq e^{-(E_1 - \delta)n_1},
\\
&\P_2 \left( A_1^{\tau} \right)
\leq e^{-(E_2 - \delta)n_2},
\end{align} 
as considered in~\cite{PolyanskiyITA2011}. In contrast, only the case where $n_1 = n_2$ is considered in~\cite{Csiszar04}. Our definition which includes the definition of~\cite{Csiszar04} as a special case is more stringent than the definition considered in~\cite{PolyanskiyITA2011}. For instance this definition does not admit sequential tests that increase the error exponent $E_1$ arbitrarily under $H_1$ by taking arbitrarily large number of samples under $H_1$ than under $H_2$, i.e., by making $\frac{n_1}{n_2}$ arbitrarily large.
\end{remarks}

In summary, an optimal fixed-length hypothesis test can only achieve the maximum error exponent in one type of error if the probability of the other error-type is kept fixed. In contrast, a sequential hypothesis test achieves both optimal error exponents simultaneously. Example~\ref{ex:al_ht} along with Fig.~\ref{fig:classical_HT} illustrates this.

\subsection{Almost-Fixed-Length Hypothesis Tests} 

We introduce a new class of hypothesis tests in the same spirit as the class of AFLF codes.

\begin{definition}
The error exponents $(E_1, E_2)$ are said to be achievable in a $(\gamma,K)$-almost-fixed-length manner for some  $\gamma \geq 0$ and $K \in \mbb{N}$, if for every $\delta > 0$ there exists an $N(\delta) \in \mbb{N}$ such that for all $n \geq N(\delta)$ there exists a hypothesis test satisfying the following
\begin{align}
&\P_i \left( \tau > n \right) \leq e^{-\gamma n} \quad  i \in \{ 1, 2\},
\\
& \tau \leq K n \quad \P_i-\mbox{a.s. for } i \in \{ 1, 2\},
\\
&\P_1 \left( A_2^{\tau} \right)
\leq e^{-(E_1 - \delta)n},
\\
&\P_2 \left( A_1^{\tau} \right)
\leq e^{-(E_2 - \delta)n}.
\end{align}

\end{definition}

Let $\mathcal{R}_{\text{AFL}}^{(\gamma, K)}$ denote the region of all feasible error exponents of the class of $(\gamma, K)$-almost-fixed-length hypothesis tests. 

\begin{remarks}
Note that as $\gamma$ tends to $\infty$, this class of tests recover the class of fixed-length hypothesis tests, hence $\mathcal{R}_{\text{FL}} = \mathcal{R}_{\text{AFL}}^{(\infty, 1)} \subset\mathcal{R}_{\text{AFL}}^{(\gamma, K)}$, for every $\gamma \geq 0$ and $K \in \mbb{N}$. Similarly, for all $\epsilon > 0$ and $n$ large enough, we have that $\expe_i[\tau] \leq n + \epsilon$, for $i \in \{1,2\}$. This implies that $\mathcal{R}_{\text{AFL}}^{(\gamma, K)} \subset \mathcal{R}_{\text{seq}}$.
\end{remarks}

\section{Error Exponents of Almost-Fixed-Length Hypothesis Tests}

Next we provide our main result for almost-fixed-length hypothesis tests.

\begin{theorem}
\label{thm:gamma_opt}
For any $\gamma \geq 0$, define 
\begin{align}
\mathcal{R}_{\gamma}
:=
\mathcal{R}_{\text{FL}}
\cup
\{E_1: E_1 \leq  E_1(\gamma)\} \times \{ E_2: E_2 \leq E_2(\gamma)\},
\end{align}
where define
\begin{align}
\label{eq:E_1_gamma}
E_1(\gamma) := \max_{\lambda \in [0,1]} \left\{ \kl{\P^{(\lambda)}}{\P_1} : \kl{\P^{(\lambda)}}{\P_2} \geq \gamma 
\right\},
\end{align}
and
\begin{align}
\label{eq:E_2_gamma}
E_2(\gamma) := \max_{\lambda \in [0,1]} \left\{ \kl{\P^{(\lambda)}}{\P_2} : \kl{\P^{(\lambda)}}{\P_1} \geq \gamma 
\right\}.
\end{align}
For any $\gamma \geq 0$ and $K \in \mbb{N}$, we have
\begin{align}
\mathcal{R}_{\gamma} \cap K\mathcal{R}_{\text{FL}} \subset \mathcal{R}_{\text{AFL}}^{(\gamma, K)}.
\end{align}
and conversely, we have 
\begin{align}
\mathcal{R}_{\text{AFL}}^{(\gamma, K)}
\subset
\mathcal{R}_{\gamma} \cap K\mathcal{R}_{\text{FL}}.
\end{align}
\end{theorem}

The proof of the above theorem follows by combining Proposition~\ref{prop:HT_2phase} with Proposition~\ref{prop:AFL_HT_converse}. The following corollary is an immediate consequence of the above theorem.

\begin{corollary}
For $\gamma > D^*$, and for all $K \in \mbb{N}$, we have \begin{align}
    \mathcal{R}_{\text{AFL}}^{(\gamma, K)} = \mathcal{R}_{\text{FL}},
\end{align}
since $\add{\{E_1: E_1 \leq  E_1(\gamma)\} \times \{E_2: E_2 \leq E_2(\gamma)\}} \subset \mathcal{R}_{\text{FL}}$.

\end{corollary}

\subsection{Achievability: A Two-Phase Hypothesis Test}
For $\gamma > D^{\ast}$, the achievability of $\mathcal{R}_{\gamma}$ coincides with that of the class of fixed-length hypothesis tests, $\mathcal{R}_{FL}$, ( $\P_i(\tau > n) = 0$ and since $\mathcal{R}_{\gamma} = \mathcal{R}_{\text{FL}}$), so any fixed-length hypothesis test achieves $\mathcal{R}_{\gamma}$. Now, let us consider $\gamma\leq D^{\ast}$ and $K \in \mathbb{N}$. We propose a hypothesis test that decides between the hypotheses at two evaluation points, one at $n$ and the other at $K n$. Formally the two phase hypothesis test is described as follows for $\gamma \leq D^{\ast}$.
\begin{enumerate}[label=(\roman*)]

\item
\underline{Phase-I:} In the first phase, we collect $n$ samples and choose whether to stop and decide between the hypotheses or to proceed to Phase-II and collect extra samples. 
Define 
\begin{align}
\label{eq:lambda_1}
&\lambda_1(\gamma)
\\
&: = 
\argmax_{\lambda \in [0,1]} \left\{\kl{\P^{(\lambda)}}{\P_1} : \kl{\P^{(\lambda)}}{\P_2} \geq \gamma \right\},
\end{align}
and 
\begin{align}
\label{eq:lambda_2}
&\lambda_2(\gamma)
\\
&: = 
\argmax_{\lambda \in [0,1]} \left\{\kl{\P^{(\lambda)}}{\P_2} : \kl{\P^{(\lambda)}}{\P_1} \geq \gamma \right\}.
\end{align}
Note that $\lambda_1(\gamma)$ and $\lambda_2(\gamma)$ denote the two $\lambda$s which achieve in the maximum in equation~\eqref{eq:E_1_gamma} and \eqref{eq:E_2_gamma} respectively. For the ease of exposition we will denote $\lambda_i(\gamma)$ as $\lambda_i$ and drop the dependence on $\gamma$ whenever it is clear from the context. Let 
\begin{align}
\label{eq:alpha_1}
\alpha_1 = \kl{\P^{(\lambda_2)}}{\P_2} - \kl{\P^{(\lambda_2)}}{\P_1},
\end{align}
\begin{align}
\label{eq:beta_1}
\beta_1 = \kl{\P^{(\lambda_1)}}{\P_2} - \kl{\P^{(\lambda_1)}}{\P_1}.
\end{align}



Phase-I of the strategy is conducted as follows: if
\begin{align}
\begin{array}{ll}
\frac{1}{n}\sum_{i = 1}^{n} \log \frac{\P_1(X_i)}{\P_2(X_i)} \geq \alpha_1 & \text{stop and choose 1},\\
\frac{1}{n}\sum_{i = 1}^{n} \log \frac{\P_1(X_i)}{\P_2(X_i)} \leq \beta_1 & \text{stop and choose 2},\\
\beta_1 < \frac{1}{n}\sum_{i = 1}^{n} \log \frac{\P_1(X_i)}{\P_2(X_i)} < \alpha_1 & \text{proceed to Phase-II}.
\end{array}
\label{eq:HT_phase1}
\end{align}

\item
\underline{Phase-II:} In the second phase, we collect $(K-1)n$ additional samples. Fix some $\lambda \in [0,1]$ and let
\begin{align}
\alpha = \kl{\P^{(\lambda)}}{\P_2} - \kl{\P^{(\lambda)}}{\P_1}. 
\end{align}
Phase-II of the strategy is conducted as follows: if 
\begin{align}
\begin{array}{ll}
\frac{1}{K n}\sum_{i =  1}^{Kn} \log \frac{\P_1(X_i)}{\P_2(X_i)} \geq \alpha & \text{stop and choose 1},\\
\frac{1}{K n}\sum_{i = 1}^{K n} \log \frac{\P_1(X_i)}{\P_2(X_i)} < \alpha & \text{stop and choose 2}.
\end{array}
\label{eq:HT_phase2}
\end{align}

\end{enumerate}

For a given hypothesis testing problem, the implementation of the above two-phase hypothesis test requires the parameters: $\gamma >0$, $K \in \mbb{N}$, $\ell \in \mbb{N}$ and $\lambda \in [0,1]$. 

\begin{proposition}
\label{prop:HT_2phase}
Let $\gamma \leq D^{\ast}$ and $K \in \mathbb{N}$. The two-phase hypothesis test as given by equations~\eqref{eq:HT_phase1} and~\eqref{eq:HT_phase2} is an almost-fixed-length hypothesis test and as $\lambda$ varies over $[0,1]$, two-phase hypothesis test achieves the error exponents on the boundary of the region 
\begin{align}
\mathcal{R}_{\gamma} \cap K \mathcal{R}_{\text{FL}}.
\end{align}
Hence, $\mc{R}_{\gamma} \cap K \mc{R}_{\text{FL}} \subset \mc{R}^{(\gamma, K)}_{\text{AFL}}$. Define
\begin{align}
    K^{\ast} := \max \left \{ \frac{\kl{\P_2}{\P_1}}{D^*}, \frac{ \kl{\P_1}{\P_2}}{D^*} \right \}.
\end{align}
Furthermore, for all $K \geq K^{\ast}$ and for $\alpha = 0$, i.e., $\lambda = \lambda^{\ast}$ in Phase-II in equation~\eqref{eq:HT_phase2}, the two phase hypothesis test achieves any $(E_1, E_2)$ on the boundary of the region $\mathcal{R}_{\gamma}$.
\end{proposition}

The proof of the above proposition is provided in Appendix~\ref{proof:HT_2phase}.

\begin{figure}[!htb]
    \includegraphics[width=0.45\textwidth]{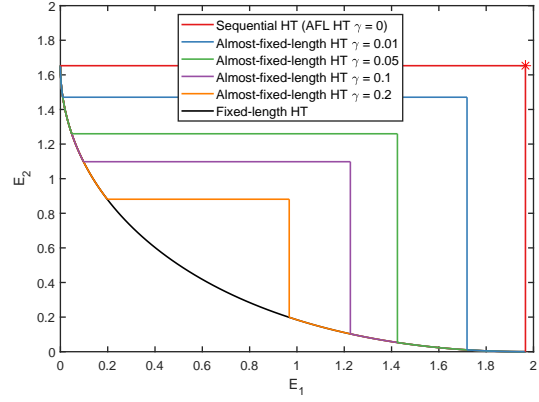}
    \caption{This figure shows the region $\mathcal{R}^{(\gamma, K)}_{\text{AFL}}$ for various values of $\gamma$ and $K \geq K^{\ast}= 4$ when the samples are Bernoulli with parameters $p_1 = 0.9$ under $H_1$ and $p_2 = 0.2$ under $H_2$. As $\gamma$ decreases the trade-off between the error exponents gets better and the test achieves the optimal sequential exponents $(\kl{\P_2}{\P_1}, \kl{\P_1}{\P_2})$.}
\label{fig:err_exp_region}
\end{figure}

\begin{figure}[!htb]
    \includegraphics[width=0.45\textwidth]{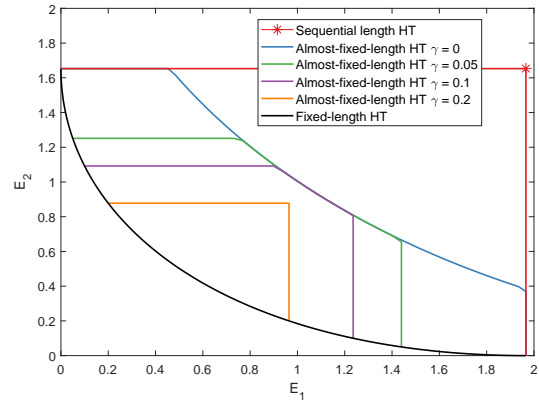}
    \caption{This figure shows the achievable region of the two phase hypothesis test as $\gamma$ increases for $K = 2$ ($K^* = 4$), when the samples are Bernoulli with parameters $p_1 = 0.9$ under $H_1$ and $p_2 = 0.2$ under $H_2$.}
\label{fig:k_2}
\end{figure}

\noindent
\textbf{Example 2 (Revisited).}
Fig.~\ref{fig:err_exp_region} shows the region of error exponents $\mathcal{R}^{(\gamma,K)}_{\text{ALF}}$ described in Theorem~\ref{thm:gamma_opt} at different values of $\gamma$ for $K \geq K^{\ast} = 4$. As $\gamma$ decreases, the trade-off between error exponents $(E_1, E_2)$ improves. In particular, it shows that it is possible to achieve the error exponents that are arbitrarily close to optimal error exponents of sequential hypothesis tests, i.e. $(\kl{\P_2}{\P_1}, \kl{\P_1}{\P_2})$, by selecting $\gamma$ arbitrarily close to zero. Fig.~\ref{fig:k_2} shows $\mc{R}_{\text{ALF}}^{(\gamma, K)}$ for $K = 2$ which is strictly less than $K^{\ast} = 4$. We see that for smaller values of $\gamma$ the feasible region of error exponents is bounded by $K\mc{R}_{\text{FL}}$.

\subsection{Converse: Hypothesis Testing with Rejection Option}

Our converse upper bounds the achievable error exponent region of a $(\gamma, K)$-almost-fixed-length hypothesis test with that of a fixed-length hypothesis test with an  additional rejection option such that the probability of rejecting a hypothesis  approaches zero exponentially fast with an exponent at most $\gamma$. After collecting $\tau$ samples, a hypothesis test with rejection option divides the sample space $\mathcal{X}^{\tau}$ into three sets or decision regions, given by $A_i^{\tau}$ for $i \in \{1, 2\}$ where the test accepts $H_i$, and $A_{\Omega}^{\tau}$ which denotes the region where the test rejects both hypotheses $H_1$ and $H_2$. Given a $(\gamma, K)$-almost-fixed-length hypothesis test we can construct a hypothesis test with rejection option by rejecting to choose either of the hypotheses whenever $\tau > n$.

\begin{definition}
The exponents $(E_1, E_2, E_{\Omega})$ are said be achievable, if for every $\delta > 0$ there exists an $N(\delta) \in \mbb{N}$ such that for all $n \geq N(\delta)$ there exists a hypothesis test with rejection option that satisfies the following
\begin{align}
&\tau \leq n \quad \P_i-\mbox{a.s. for } i \in \{ 1, 2\},
\\
&\P_1(A_2^{\tau}) \leq e^{-(E_1 -\delta)n}, \quad 
\P_2(A_1^{\tau}) \leq e^{-(E_2 -\delta)n},
\\
&\P_1(A_{\Omega}^{\tau}) + \P_2(A_{\Omega}^{\tau}) \leq e^{-(E_{\Omega} -\delta)n}.
\end{align}
\end{definition}

\begin{lemma}
\label{lemma:HT_rej_region}
For any $\gamma \geq 0$, let $\mathcal{R}^{(\gamma)}_{\Omega}$ denote the region of all feasible error exponents for the class of hypothesis tests with a rejection option, then we have
\begin{align}
\mathcal{R}_{\gamma} \times \{ E_{\Omega}  = \gamma\}
\subset 
\mathcal{R}^{(\gamma)}_{\Omega}.
\end{align}
Conversely, for every $\gamma \geq 0$ we have
\begin{align}
\mathcal{R}^{(\gamma)}_{\Omega}
&\subset
\mc{R}_{\gamma} \times \{ E_{\Omega}  = \gamma\}.
\end{align}
\end{lemma}
A variant of above the lemma has been proved under the class of hypothesis tests with rejection option considered by Grigoryan et~al.~in~\cite{multi_hyp_rej}, Sason in~\cite{mod_dev_HT_Sason} and Gutman in~\cite{gutman}. The following proposition builds on the converse of hypothesis tests with a rejection option to provide a converse for almost-fixed-length hypothesis tests.


\begin{proposition}
\label{prop:AFL_HT_converse}
Let $\gamma \geq 0$ and $K \in \mathbb{N}$. The region of all feasible error exponents of the class of $(\gamma, K)$-almost-fixed-length tests satisfies
\begin{align}
\mc{R}_{\text{AFL}}^{(\gamma, K)} 
\subset
\mc{R}_{\gamma}\cap K\mc{R}_{\text{FL}}.
\end{align}
\end{proposition}

The proof of the above proposition is provided in Appendix~\ref{proof:AFL_HT_converse}.

\section{Conclusion and Future Work}

\label{sec:conclusion}
We looked at a new class of strategies for channel coding and hypothesis tests that have a slight flexibility over fixed-length strategies by allowing a slightly large stopping time on rare set of sample paths with exponentially small probability. We show that when stopping times are allowed to be slightly large in only exponentially small cases, the overall reliability is increased significantly. We showed that it is possible to achieve optimal performance of variable-length strategies using almost-fixed-length strategies. The performances achieved by the two settings, namely variable-length setting and fixed-length setting that were thought to be very distinct, are, in fact, the extremities of a continuum of performance curves achieved by almost-fixed-length coding strategies parametrized by $\gamma$.  Recall that for any $n \geq 1$, for all $(\ell, M, \gamma, K, \epsilon)$ AFLF code and $(\gamma, K)$ AFL hypothesis test we have 
\begin{align}
    \lim_{\ell \to \infty} \expe\left[\left(\frac{\tau_{\ell}}{\ell}\right)^{n}\right] = 1, \quad \lim_{\ell \to \infty}\text{Var}(\tau_{\ell}) = 0.
\end{align}
This means that the class of strategies for which the variance of the stopping time is required be to zero, is no more restrictive than the class of strategies that satisfy only an average-length constraint, in terms of reliability and error exponents. Hence, we showed that growing variability not essential to obtaining the optimal error exponents. Similar statements can be made for constraining higher moments of the stopping time.

%


\appendix

\subsection{Two-Phase AFLF Code based on Truncated Yamamoto-Itoh Strategy}

\label{app:two_phase_code}

The following achievability strategy is similar to the strategy considered in~\cite{Nakiboglu_error_erasure_feedback}, however for completeness we provide a simpler and intuitive proof which is more natural to our problem. Our two-phase strategy is described as follows:
\begin{enumerate}
\item
\underline{Phase-I (Truncated Yamamoto-Itoh Strategy):} Fix some $\alpha \in [0,1]$. For the first phase we consider the Yamamoto-Itoh strategy~\cite{YI_1979} with block-length $\ell$ to transmit $2^{\ell R}$ number of messages. The first phase is divided into two parts, where each part of length $\alpha \ell$ and $\left( 1 - \alpha \right)\ell$. In the first part, we transmit $\ell R$ bits over $\alpha \ell$ channel uses using a random code~\cite{Gallager_book}. Hence, the probability of making an error in decoding the transmitted message in the first part can be upper bounded as
\begin{align}
\P_{\text{1e}} \leq 2^{-\alpha \ell E_{\text{r}}\left(\frac{R}{\alpha}\right)}.
\end{align}
Recall that $x, x^{\prime} \in \mc{X}$ channel inputs for which $D(P_{Y|X}(\cdot|x)||\P_{Y|X}(\cdot|x^{\prime}))$ is maximized, i.e., $C_1 = D(P_{Y|X}(\cdot|x)||\P_{Y|X}(\cdot|x^{\prime}))$. If the received message has been decoded correctly the encoder send ACKs\footnote{The encoder has access to the decoded message at the decoder because of the noiseless and delay-free feedback channel.} by transmitting channel input $x$ otherwise we send NACKs by transmitting channel input $x^{\prime}$ for the remaining $\left( 1 - \alpha \right)\ell$ channel uses.
Now construct a fixed-length hypothesis test as shown in Fact~\ref{fact:fl_ht}, for some $\lambda \in [0,1]$, to distinguish between the ACK and NACK symbols received such that 
\begin{align}
\P_{\text{2ec}} \leq 2^{-\left( 1 - \alpha \right)\ell D(\P^{(\lambda)}||\P_{Y|X}(\cdot|x^{\prime}))},
\end{align}
and
\begin{align}
\P_{\text{2ce}} \leq 2^{-\left( 1 - \alpha \right)\ell D(\P^{(\lambda)}||\P_{Y|X}(\cdot|x))},
\end{align}
where $\P_{\text{2ec}}$ denotes the probability an ACK is decoded when NACK was transmitted and $\P_{\text{2ce}}$ denotes the probability a NACK is decoded when ACK was transmitted.
If the hypothesis test declares a NACK, proceed to Phase-II otherwise accept the decoded message.

\item
\underline{Phase-II (Random Code):} In the second phase send $\ell R$ bits over $(K-1)\ell$ channel uses using a random code at effective rate of $\frac{R}{K-1}$ so that $\tau \leq K\ell$. 

\end{enumerate}

For a given channel and rate, the implementation of the above two-phase channel code requires the parameters: $\alpha >0$, $K \in \mbb{N}$, $\ell \in \mbb{N}$ and $\lambda \in [0,1]$. 

Next, we show that the stopping time $\tau$ of the above strategy satisfies the constraints of an AFLF code's stopping time. By construction, length of Phase-I is $\ell$ and length of Phase-II is $(K-1)\ell$. Hence, the stopping time is almost surely bounded by $K\ell$. The probability of the stopping time exceeding $\ell$ equivalently the probability of entering Phase-II is given by 
\begin{align}
\P(\tau > \ell) 
&= \P_{1e} (1-\P_{\text{2ec}}) + (1-\P_{1e})\P_{\text{2ce}} 
\\
&\leq \P_{\text{1e}} + \P_{\text{2ce}}
\\
&\leq 2^{- \alpha  \ell E_{\text{r}}\left(\frac{R}{\alpha}\right)} + 2^{-\left( 1 - \alpha \right)\ell D(\P^{(\lambda)}||\P_{Y|X}(\cdot|x))}
\\
& \leq 2.2^{-\min\left\{\alpha E_{\text{r}}\left(\frac{R}{\alpha}\right), \left( 1 - \alpha \right) D(\P^{(\lambda)}||\P_{Y|X}(\cdot|x)) \right\} \ell}.
\end{align}
This implies the stopping time $\tau$ of two-phase channel code exceeds $\ell$ with an exponentially small probability with an exponent $\gamma$ equal to $ \min \left\{\alpha E_{\text{r}}\left(\frac{R}{\alpha}\right), \left( 1 - \alpha \right) D(\P^{(\lambda)}||\P_{Y|X}(\cdot|x))\right\}$. Hence, this two-phase channel code belongs to the class of AFLF codes.

Furthermore, let $\epsilon_{(K-1)\ell}$ denote the probability of error in Phase-II then the total probability of error of the two-phase channel code is given by
\begin{align}
\epsilon_{\ell}
&=
\P_{\text{1e}}\P_{\text{2ec}} + \P(\tau > \ell)\epsilon_{(K-1)\ell}
\nonumber
\\
&\leq 
2^{-\alpha \ell E_{\text{r}}\left(\frac{R}{\alpha}\right)}2^{-\left( 1 - \alpha \right)\ell D(\P^{(\lambda)}||\P_{Y|X}(\cdot|x^{\prime}))} 
\\
&\quad + 2^{-(K-1)\ell E_{\text{r}} \left(\frac{R}{K-1}\right)}
\\
& \leq 2. 2^{-\left\{ \alpha  E_{\text{r}}\left(\frac{R}{\alpha}\right) +\left( 1 - \alpha \right)D(\P^{(\lambda)}||\P_{Y|X}(\cdot|x^{\prime})), (K-1)E_{\text{r}} \left(\frac{R}{K-1}\right)\right\} \ell}.
\end{align}

\subsection{Proof of Theorem~\ref{thm:achv_aflf_gamma_0}}

\label{proof:achv_aflf_gamma_0}

Consider the two-phase channel code proposed in Appendix~\ref{app:two_phase_code}. Fix $\epsilon> 0$ and let $\alpha = \frac{R}{C-\epsilon}$. We construct the hypothesis test in Phase-I such that the exponent of $\P_{2ce}$ error is $0$ and the exponent of $\P_{2ec}$ is $D(\P_{Y|X}(\cdot|x)||\P_{Y|X}(\cdot|x^{\prime}))$, i.e., $C_1$ of the channel. For that we choose $\lambda = 0$ so that $\P^{\lambda}(y) = \P_{Y|X}( y|x)$ for all $y$. Hence, exponential rate of the probability of entering Phase-II is $0$, i.e., exponential rate of stopping time $\tau$ exceeding $\ell$ is $0$, which implies $\gamma = 0$. Now, as $\epsilon$ goes to $0$ we have $\alpha$ goes to $\frac{R}{C}$, this implies $\alpha E_r(R/\alpha)$ goes to $0$. Therefore, for $\lambda = 0$ and as $\epsilon$ goes to $0$, and the error exponent of the two-phase channel code satisfies
\begin{align}
&\liminf_{\ell \to \infty}-\frac{1}{\ell}\log \epsilon_{\ell} 
\\
&=
\min \left\{ C_1 \left( 1 - \frac{R}{C}\right), 
(K-1) E_{\text{r}} \left(\frac{R}{K-1}\right)
\right \}.
\end{align}

To obtain the value of $K$ for which the Burnashev's optimal $\mc{E}_{\text{VLF}}(R)$ is achieved using the above two-phase channel code, we apply the argument used in~\cite{arq_erasure_gopala2007} for any general DMC. Since $C_1 > C$ we have
\begin{align}
\frac{\partial \mc{E}_{\text{VLF}}(R)}{\partial R} = -\frac{C_1}{C} \leq -1.
\end{align}
Furthermore, it is known that~\cite{arq_erasure_gopala2007}
\begin{align}
\frac{\partial (K-1) E_{\text{r}}(\frac{R}{K-1})}{\partial R} \geq -1.
\end{align}
This implies the for any rate $R$, the rate of decrease of the Burnashev exponent $\mc{E}_{\text{VLF}}$ is higher than that of $(K-1)E_{\text{r}}(\frac{R}{K-1})$. Now, we have 
\begin{align}
(K-1)E_{\text{r}}\left(\frac{C}{K-1}\right) \geq \mc{E}_{\text{VLF}}(C) = 0, 
\end{align}
and if $K \geq 1+\frac{C_1}{E_{\text{r}}(0)}$, then at $R = 0$, we have
\begin{align}
(K-1)E_{\text{r}}\left(0\right) \geq \mc{E}_{\text{VLF}}(0) = C_1. 
\end{align}
Hence, the curve $(K-1)E_{\text{r}}\left(\frac{R}{K-1}\right)$ lies strictly above the curve $\mc{E}_{\text{VLF}}(R)$ for all $0 < R< C$, which implies the error exponent of the two-phase code achieves Burnashev's optimal exponent $\mc{E}_{\text{VLF}}$ when $K \geq K^{\ast}:=  1+\frac{C_1}{E_{\text{r}}(0)}$. Combining with Corollary~\ref{coro:aflf_basic_bounds} we have the assertion of the theorem.

\subsection{Proof of Theorem~\ref{thm:achv_aflf_any_gamma}}
\label{proof:achv_aflf_any_gamma}

Fix $\alpha, \lambda \in [0,1]$ in the two-phase channel code described in Appendix~\ref{app:two_phase_code}. Now, note that
\begin{align}
&\liminf_{\ell \to \infty} -\frac{1}{\ell} \log \P(\tau > \ell)
\\
&\geq 
\min\left\{\alpha   E_{\text{r}}\left(\frac{R}{\alpha}\right), \left( 1 - \alpha \right) D(\P^{(\lambda)}||\P_{Y|X}(\cdot|x))\right\},
\end{align}
and similarly we have
\begin{align}
&\mc{E}_{\text{AFLF}}(R,\gamma, K)
\\
&\geq 
\min\left\{\alpha   E_{\text{r}}\left(\frac{R}{\alpha}\right)+ \left( 1 - \alpha \right) D(\P^{(\lambda)}||\P_{Y|X}(\cdot|x^{\prime})), \right.
\\
& \left. \hspace{2cm}(K-1)E_{\text{r}}\left( \frac{R}{K-1}\right)\right\}.
\end{align}

Next, for a given value of $\gamma$ and a rate $R$ below capacity, we show how to set $\alpha$ and $\lambda$ values such that $\P(\tau > \ell) < 2^{-\gamma \ell}$ and also obtain the error exponent achieved by the two-phase channel code.

\underline{Case 1:} Consider the case where $0 < \gamma < E(R)$. Let $\alpha^{\ast}(R, \gamma)$ denote $\alpha$ which satisfies
\begin{align}
\alpha E_{\text{r}}\left(\frac{R}{\alpha} \right) = \gamma,
\end{align}
and hence we can write
\begin{align}
\alpha^{\ast}(R, \gamma) = \frac{R}{g^{-1}(\frac{\gamma}{R})},
\end{align}
where $g(a) = \frac{E_{\text{r}}(a)}{a}$. Then, we have 
\begin{align}
\mc{E}_{\text{AFLF}}(R, \gamma, K) 
\geq
\min \left\{
E_{\text{ee}}^{\prime}(R, \gamma), (K-1)E_{\text{r}}\left( \frac{R}{K-1}\right)
\right\} ,
\end{align}
where we define
\begin{align}
&E_{\text{ee}}^{\prime}(R, \gamma) 
: = 
\\
&\max_{\alpha \in [\alpha^{\ast}(R, \gamma), 1]} 
\max_{\substack{
\lambda \in [0,1] 
\\  \left( 1 - \alpha \right) D(\P^{(\lambda)}||\P_{Y|X}(\cdot|x))
\geq \gamma}}E_{\text{ee}}^{\prime\prime}(\alpha, \lambda, R),
\label{eq:alpha_lambda}
\end{align}
and define
\begin{align}
E_{\text{ee}}^{\prime\prime}(\alpha, \lambda, R)
: =  \alpha   E_{\text{r}}\left(\frac{R}{\alpha}\right)+ \left( 1 - \alpha \right) D(\P^{(\lambda)}||\P_{Y|X}(\cdot|x^{\prime})).
\end{align}
This implies we choose $\alpha$ and $\lambda$ values which attain the maximum in equation~\eqref{eq:alpha_lambda}.

\underline{Case 2:} Consider the case where $\gamma > E_{\text{r}}(R)$. Since for all $\alpha \in [0,1]$ we have $\alpha E_{\text{r}}\left(\frac{R}{\alpha} \right) \leq E_{\text{r}}(R)$, choose $\alpha = 1$ so that the decoder never declares NACK and hence $\tau = \ell$ a.s. Hence, we have $\mc{E}_{\text{AFLF}}(R,\gamma, K) \geq E_{r}(R)$ for all $0 \leq R \leq C$.

\subsection{Proof of Proposition~\ref{prop:converse}}
\label{proof:converse}

For every $(\ell, M, \gamma, K, \epsilon)$ AFLF code we construct an $(\ell, M, \epsilon, 2^{-\gamma \ell})$ fixed-length error-erasure feedback code as follows. Using an $(\ell, M, \gamma, K, \epsilon)$ AFLF code, after $\ell$ channel uses if $\tau > \ell$ then declare an erasure. Since $\P(\tau > \ell) < 2^{-\gamma \ell}$, this implies the probability of erasure is at most $2^{-\gamma \ell}$. In other words, the erasure exponent of this code is $\gamma$. Since the optimal error exponent of a fixed-length error-erasure feedback code with erasure exponent $\gamma$ is given by $\mc{E}_{\text{ee}}(R, \gamma)$, the probability of error of $(\ell, M, \epsilon, 2^{-\gamma \ell})$ error-erasure code satisfies
\begin{align}
\epsilon \geq 2^{-\ell \mc{E}_{\text{ee}}(R, \gamma)}.
\end{align}
Therefore, the probability of error of any $(\ell, M, \gamma, K, \epsilon)$ AFLF code must be at least $2^{-\ell \mc{E}_{\text{ee}}(R, \gamma)}$, i.e., $\mc{E}_{\text{AFLF}}(R, \gamma, K) \leq \mc{E}_{\text{ee}}(R, \gamma)$. Furthermore, the AFLF code uses $K\ell$ channel uses which implies probability of error must be at least equal to the error of a fixed-length feedback code with code-length $K \ell$ with rate $\frac{R}{K}$. This implies
\begin{align}
\epsilon \geq 2^{-K E_{\text{H}}(\frac{R}{K})},
\quad 
\text{and}
\quad
\mc{E}_{\text{AFLF}}(R, \gamma, K) \leq K E_{\text{H}}\left(\frac{R}{K}\right).
\end{align}
In other words, the probability of error of $(\ell, M, \gamma, K, \epsilon)$ AFLF code
\begin{align}
\epsilon \geq \max\left\{ 2^{-\ell E_{\text{ee}}(R, \gamma)}, 2^{-K E_{\text{H}}\left(\frac{R}{K}\right)}\right\},
\end{align}
and hence we have $\mc{E}_{\text{AFLF}}(R, \gamma, K) \leq \min\{\mc{E}_{\text{ee}}(R, \gamma), K E_{\text{H}}(\frac{R}{K})\}$

\subsection{Proof of Proposition~\ref{prop:HT_2phase}}

\label{proof:HT_2phase}
It is straightforward to check that equations~\eqref{eq:lambda_1} and \eqref{eq:lambda_2} imply
\begin{align}
\label{eq:lambda_1_implication}
&D\left( \P^{(\lambda_1)}||\P_1\right) = E_1(\gamma), \text{ and } 
D\left( \P^{(\lambda_1)}||\P_2\right) = \gamma,
\\
&D\left( \P^{(\lambda_2)}||\P_2\right) = E_2(\gamma), \text{ and } 
D\left( \P^{(\lambda_2)}||\P_1\right) = \gamma.
\label{eq:lambda_2_implication}
\end{align}

Note that by construction for the two-phase hypothesis test we have $\tau \leq K n$ a.s. Additionally, we must show that the probability of the sample paths where $\tau > n$ is exponentially small with an exponent $\gamma$. Consider
\begin{align*}
\P_1(\tau > n) 
&=
\P_1\left( \beta_1 < \frac{1}{ n}\sum_{i = 1}^{n} \log \frac{\P_1(X_i)}{\P_2(X_i)} < \alpha_1 \right) 
\nonumber
\\
&\leq 
\P_1\left( \frac{1}{ n}\sum_{i = 1}^{n} \log \frac{\P_1(X_i)}{\P_2(X_i)} < \alpha_1 \right).
\end{align*}
Hence, for any $\delta > 0$ there exists an $N(\delta)$ such that for all $n \geq N(\delta)$ we have
\begin{align*}
\P_1(\tau > n) 
\overset{(a)}\leq e^{-(\kl{\P^{(\lambda_2)}}{\P_1}-\delta) n}
 \overset{(b)}\leq e^{-(\gamma-\delta) n},
\end{align*}
where $(a)$ is obtained using Sanov's Theorem (Theorem 11.4.1 in~\cite{CoverBook2nd}) and equation~\eqref{eq:beta_1}, and $(b)$ follows from equation~\eqref{eq:lambda_2_implication}. Similarly, for all $n \geq N(\delta)$ we also have $\P_2(\tau > n) \leq e^{-(\gamma-\delta) n}$ using Sanov's Theorem and equations~\eqref{eq:alpha_1} and \eqref{eq:lambda_1_implication}. Hence, this test belongs to the class of $(\gamma, K)$-almost-fixed-length hypothesis test.

The error of type-I is given as follows,
\begin{align*}
&\P_1\left( A_2^{\tau}\right)
=
\P_1 \left( \frac{1}{n}\sum_{i = 1}^{n} \log \frac{\P_1(X_i)}{\P_2(X_i)} \leq \beta_1 \right)
+ 
\\
&
\hspace{0.5cm}\P_1 \left( \left\{\beta_1 < \frac{1}{n}\sum_{i = 1}^{n} \log \frac{\P_1(X_i)}{\P_2(X_i)} < \alpha_1 \right.\right\}
\\
&
\hspace{0.5cm} \cap \left. \left\{\frac{1}{K n}\sum_{i = 1}^{K n} \log \frac{\P_1(X_i)}{\P_2(X_i)} < \alpha \right\}\right).
\end{align*}
Fix $\lambda \in [0,1]$. Using Sanov's Theorem and from the definition of $\alpha_1$ and $\beta_1$, for any $\delta> 0$ there exists an $N_0(\delta)$ such that for all $n \geq N_0(\delta)$ we have
\begin{align*}
&\P_1\left( A_2^{\tau}\right)
\leq
e^{-(\kl{\P^{(\lambda_1)}}{\P_1}-\delta) n} + e^{-(K\kl{\P^{(\lambda)}}{\P_1} -\delta) n}.
\end{align*}
Now, taking limit we obtain
\begin{align}
&\lim_{n \to \infty}\frac{1}{n} -\log \P_1 \left( A_2^{\tau}\right)
\\
&\geq
\min \left\{  \kl{\P^{(\lambda_1)}}{\P_1},  K\kl{\P^{(\lambda)}}{\P_1}\right\}
\\
& = \min \{E_1(\gamma), K\kl{\P^{(\lambda)}}{\P_1}\}.
\label{eq:exp1}
\end{align}
Similarly, we obtain
\begin{align}
&\lim_{n \to \infty}\frac{1}{n} -\log \P_2 \left( A_1^{\tau}\right)
\\
&\geq
\min \left\{  \kl{\P^{(\lambda_2)}}{\P_2},  K\kl{\P^{(\lambda)}}{\P_2}\right\}
\\
& = \min\{E_2(\gamma), K\kl{\P^{(\lambda)}}{\P_2}\}.
\label{eq:exp2}
\end{align}
Equations~\eqref{eq:exp1} and~\eqref{eq:exp2} imply that by changing the value of $\lambda$ over the set $[0,1]$, the two-phase hypothesis test can achieve the error exponents on the boundary of the region $\mc{R}_{\gamma} \cap K\mc{R}_{\text{FL}}$. Now if we set $\alpha = 0$ in Phase-II, i.e., set $\lambda = \lambda^{\ast}$, then we have 
$\kl{\P^{(\lambda)}}{\P_1} = \kl{\P^{(\lambda)}}{\P_2} = D^{\ast}$. Hence, we have the assertion of the proposition.

\subsection{Proof of Proposition~\ref{prop:AFL_HT_converse}}
\label{proof:AFL_HT_converse}

From the definition of $(\gamma, K)$-almost-fixed-length hypothesis tests we have
\begin{align}
\mc{R}_{\text{AFL}}^{(\gamma, K)}
\subset
\mc{R}_{\text{AFL}}^{(0, K)}
\subset
K\mc{R}_{\text{FL}}.
\end{align}
Given a $(\gamma, K)$-almost-fixed-length hypothesis test we can construct a hypothesis test with rejection option by rejecting to choose either of the hypotheses whenever $\tau > n$. From the converse of hypothesis test with a rejection option in Lemma~\ref{lemma:HT_rej_region}, we have
\begin{align}
\mc{R}_{\text{AFL}}^{(\gamma, K)}\times \{E_{\Omega} = \gamma\}
\subset
\mc{R}_{\gamma}\times \{E_{\Omega} = \gamma\}.
\end{align}
Therefore, for every $\gamma \geq 0$ and $K \in \mbb{N}$ we have 
\begin{align}
\mc{R}_{\text{AFL}}^{(\gamma, K)}
\subset
\mc{R}_{\gamma} \cap K\mc{R}_{\text{FL}}.
\end{align}
Hence, we have the converse for Theorem~\ref{thm:gamma_opt}.


\bibliographystyle{IEEEtran}
\bibliography{HypTest}

\begin{thebibliography}{10}
\providecommand{\url}[1]{#1}
\csname url@samestyle\endcsname
\providecommand{\newblock}{\relax}
\providecommand{\bibinfo}[2]{#2}
\providecommand{\BIBentrySTDinterwordspacing}{\spaceskip=0pt\relax}
\providecommand{\BIBentryALTinterwordstretchfactor}{4}
\providecommand{\BIBentryALTinterwordspacing}{\spaceskip=\fontdimen2\font plus
\BIBentryALTinterwordstretchfactor\fontdimen3\font minus
  \fontdimen4\font\relax}
\providecommand{\BIBforeignlanguage}[2]{{%
\expandafter\ifx\csname l@#1\endcsname\relax
\typeout{** WARNING: IEEEtran.bst: No hyphenation pattern has been}%
\typeout{** loaded for the language `#1'. Using the pattern for}%
\typeout{** the default language instead.}%
\else
\language=\csname l@#1\endcsname
\fi
#2}}
\providecommand{\BIBdecl}{\relax}
\BIBdecl

\bibitem{7541591}
A.~{Lalitha} and T.~{Javidi}, ``Reliability of sequential hypothesis testing
  can be achieved by an almost-fixed-length test,'' in \emph{2016 IEEE
  International Symposium on Information Theory (ISIT)}, 2016, pp. 1710--1714.

\bibitem{CoverBook2nd}
T.~M. Cover and J.~A. Thomas, \emph{Elements of information theory (2nd
  ed)}.\hskip 1em plus 0.5em minus 0.4em\relax New York, NY, USA: John Wiley \&
  Sons, Inc., 2006.

\bibitem{Burnashev76}
\BIBentryALTinterwordspacing
M.~V. Burnashev, ``{Data transmission over a discrete channel with feedback.
  Random transmission time},'' \emph{Problemy Peredachi Informatsii}, vol.~12,
  no.~4, pp. 10--30, 1976. [Online]. Available:
  \url{http://mi.mathnet.ru/ppi1706}
\BIBentrySTDinterwordspacing

\bibitem{error_exp_blahut}
\BIBentryALTinterwordspacing
R.~Blahut, ``Hypothesis testing and information theory,'' \emph{IEEE
  Transactions on Information Theory}, vol.~20, no.~4, pp. 405--417, Jul 1974.
  [Online]. Available: \url{http://dx.doi.org/10.1109/TIT.1974.1055254}
\BIBentrySTDinterwordspacing

\bibitem{Error_exp_tuncel}
\BIBentryALTinterwordspacing
E.~Tuncel, ``Extensions of error exponent analysis in hypothesis testing,'' in
  \emph{Proceedings of International Symposium on Information Theory (ISIT)},
  Sept 2005, pp. 835--839. [Online]. Available:
  \url{http://dx.doi.org/10.1109/ISIT.2005.1523454}
\BIBentrySTDinterwordspacing

\bibitem{Wald48}
\BIBentryALTinterwordspacing
A.~Wald and J.~Wolfowitz, ``Optimum character of the sequential probability
  ratio tests,'' \emph{The Annals of Mathematical Statistics}, vol.~19, no.~3,
  pp. 326--339, 1948. [Online]. Available:
  \url{http://dx.doi.org/10.1214/aoms/1177730197}
\BIBentrySTDinterwordspacing

\bibitem{Chernoff59}
\BIBentryALTinterwordspacing
H.~Chernoff, ``Sequential design of experiments,'' \emph{The Annals of
  Mathematical Statistics}, vol.~30, pp. 755--770, 1959. [Online]. Available:
  \url{https://doi.org/10.1214/aoms/1177706205}
\BIBentrySTDinterwordspacing

\bibitem{YI_1979}
\BIBentryALTinterwordspacing
H.~Yamamoto and K.~Itoh, ``Asymptotic performance of a modified
  schalkwijk-barron scheme for channels with noiseless feedback (corresp.),''
  \emph{IEEE Transactions on Information Theory}, vol.~25, no.~6, pp. 729--733,
  Nov 1979. [Online]. Available:
  \url{http://dx.doi.org/10.1109/TIT.1979.1056119}
\BIBentrySTDinterwordspacing

\bibitem{yucel_allerton_2015}
\BIBentryALTinterwordspacing
Y.~{Altuğ}, H.~V. {Poor}, and S.~{Verdú}, ``Variable-length channel codes
  with probabilistic delay guarantees,'' in \emph{Proceedings of 53rd Annual
  Allerton Conference on Communication, Control, and Computing (Allerton)},
  Sep. 2015, pp. 642--649. [Online]. Available:
  \url{http://dx.doi.org/10.1109/ALLERTON.2015.7447065}
\BIBentrySTDinterwordspacing

\bibitem{Forney1968}
\BIBentryALTinterwordspacing
G.~Forney, ``Exponential error bounds for erasure, list, and decision feedback
  schemes,'' \emph{IEEE Transactions on Information Theory}, vol.~14, no.~2,
  pp. 206--220, Mar 1968. [Online]. Available:
  \url{http://dx.doi.org/10.1109/TIT.1968.1054129}
\BIBentrySTDinterwordspacing

\bibitem{erasure_exp_telatar1994}
\BIBentryALTinterwordspacing
I.~E. Telatar and R.~G. Gallager, ``New exponential upper bounds to error and
  erasure probabilities,'' in \emph{Proceedings of IEEE International Symposium
  on Information Theory (ISIT)}, Jun 1994, pp. 379--. [Online]. Available:
  \url{http://dx.doi.org/10.1109/ISIT.1994.394640}
\BIBentrySTDinterwordspacing

\bibitem{Nakiboglu_error_erasure_feedback}
\BIBentryALTinterwordspacing
B.~Nakiboglu and L.~Zheng, ``Errors-and-erasures decoding for block codes with
  feedback,'' \emph{IEEE Transactions on Information Theory}, vol.~58, no.~1,
  pp. 24--49, Jan 2012. [Online]. Available:
  \url{http://dx.doi.org/10.1109/TIT.2011.2169529}
\BIBentrySTDinterwordspacing

\bibitem{multi_hyp_rej}
\BIBentryALTinterwordspacing
N.~Grigoryan, A.~Harutyunyan, S.~Voloshynovskiy, and O.~Koval, ``On multiple
  hypothesis testing with rejection option,'' in \emph{Proceedings of IEEE
  Information Theory Workshop (ITW)}, Oct 2011, pp. 75--79. [Online].
  Available: \url{http://dx.doi.org/10.1109/ITW.2011.6089531}
\BIBentrySTDinterwordspacing

\bibitem{mod_dev_HT_Sason}
\BIBentryALTinterwordspacing
I.~Sason, ``Moderate deviations analysis of binary hypothesis testing,'' in
  \emph{Proceedings of IEEE International Symposium on Information Theory
  (ISIT)}, July 2012, pp. 821--825. [Online]. Available:
  \url{http://dx.doi.org/10.1109/ISIT.2012.6284675}
\BIBentrySTDinterwordspacing

\bibitem{haroutunian_1977}
E.~A. Haroutunian, ``A lower bound on the probability of error for channels
  with feedback,'' vol.~13, no.~2, pp. 36--44, 1977.

\bibitem{Gallager_book}
R.~G. Gallager, \emph{Information Theory and Reliable Communication}.\hskip 1em
  plus 0.5em minus 0.4em\relax New York, NY, USA: John Wiley \& Sons, Inc.,
  1968.

\bibitem{upper_bound_nakiboglu}
\BIBentryALTinterwordspacing
B.~{Nakiboğlu} and L.~{Zheng}, ``Upper bounds to error probability with
  feedback,'' in \emph{Proceedings of 47th Annual Allerton Conference on
  Communication, Control, and Computing (Allerton)}, Sep. 2009, pp. 865--871.
  [Online]. Available: \url{http://dx.doi.org/10.1109/ALLERTON.2009.5394953}
\BIBentrySTDinterwordspacing

\bibitem{yuri_feedback_2011}
\BIBentryALTinterwordspacing
Y.~{Polyanskiy}, H.~V. {Poor}, and S.~{Verdu}, ``Feedback in the non-asymptotic
  regime,'' \emph{IEEE Transactions on Information Theory}, vol.~57, no.~8, pp.
  4903--4925, Aug 2011. [Online]. Available:
  \url{http://dx.doi.org/10.1109/TIT.2011.2158476}
\BIBentrySTDinterwordspacing

\bibitem{arq_erasure_gopala2007}
\BIBentryALTinterwordspacing
P.~K. Gopala, Y.~H. Nam, and H.~E. Gamal, ``On the error exponents of arq
  channels with deadlines,'' \emph{IEEE Transactions on Information Theory},
  vol.~53, no.~11, pp. 4265--4273, Nov 2007. [Online]. Available:
  \url{http://dx.doi.org/10.1109/TIT.2007.907431}
\BIBentrySTDinterwordspacing

\bibitem{Csiszar04}
\BIBentryALTinterwordspacing
I.~Csiszar and P.~C. Shields, ``Information theory and statistics: a
  tutorial,'' \emph{Foundations and Trends in Communications and Information
  Theory}, vol.~1, no.~4, pp. 417--528, December 2004. [Online]. Available:
  \url{http://dx.doi.org/10.1561/0100000004}
\BIBentrySTDinterwordspacing

\bibitem{PolyanskiyITA2011}
Y.~Polyanskiy and S.~Verdu, ``Binary hypothesis testing with feedback,'' in
  \emph{Information Theory and Applications Workshop (ITA)}, 2011.

\bibitem{gutman}
\BIBentryALTinterwordspacing
M.~Gutman, ``Asymptotically optimal classification for multiple tests with
  empirically observed statistics,'' \emph{IEEE Transactions on Information
  Theory}, vol.~35, no.~2, pp. 401--408, March 1989. [Online]. Available:
  \url{http://dx.doi.org/10.1109/18.32134}
\BIBentrySTDinterwordspacing

\end{thebibliography}


\end{document}